\documentclass[sigconf]{acmart}

\AtBeginDocument{%
  \providecommand\BibTeX{{%
    \normalfont B\kern-0.5em{\scshape i\kern-0.25em b}\kern-0.8em\TeX}}}

\usepackage{amssymb,amsfonts}
\usepackage{graphicx}
\usepackage{textcomp}
\usepackage{ctable}

\usepackage{algorithm}
\usepackage{algpseudocode}
\floatname{algorithm}{Protocol}

\usepackage{amsmath}

\newcommand*{\QEDB}{\hfill\ensuremath{\square}}
\newcommand{\vecv}{\vec{v}\@ifnextchar{^}{\,}{}}
\newcommand{\vect}{\vec{t}\@ifnextchar{^}{\,}{}}

\usepackage{multirow}
\usepackage{array}
\usepackage{diagbox}
\usepackage{xspace}

\newcommand{\sysname}{RevFRF\xspace}
\newcommand{\entity}[1]{#1}
\newcommand{\func}[1]{{\textsf{\small{#1}}}}
\newcommand{\protocol}[1]{\texttt{#1}}

\renewcommand\arraystretch{1.4}

\usepackage{subfigure}
\usepackage{epstopdf}

\begin{document}
\title{Revocable Federated Learning: A Benchmark of Federated Forest}

\author{Yang Liu$^1$, Zhuo Ma*$^1$, Ximeng Liu*$^2$, Zhuzhu Wang$^1$, Siqi Ma$^3$, Kui Ren$^4$}
\affiliation{%
  \institution{
  * Corresponding Author\\
  $^1$ Xidian University,
  $^2$ Fuzhou University,
  $^3$ Data 61, CSIRO,
  $^4$ Zhejiang University}
}
\email{bcds2018@foxmail.com, mazhuo@mail.xidian.edu.cn, snbnix@gmail.com} \email{z.z.wang@foxmail.com, siqi.ma@csiro.au, kuiren@zju.edu.cn}

\begin{abstract}
A learning federation is composed of multiple participants who use the federated learning technique to collaboratively train a machine learning model without directly revealing the local data.
Nevertheless, the existing federated learning frameworks have a serious defect that even a participant is revoked, its data are still remembered by the trained model.
In a company-level cooperation, allowing the remaining companies to use a trained model that contains the memories from a revoked company is obviously unacceptable, because it can lead to a big conflict of interest.
Therefore, we emphatically discuss the participant revocation problem of federated learning and design a revocable federated random forest (RF) framework, \sysname, to further illustrate the concept of revocable federated learning.
In \sysname, we first define the security problems to be resolved by a revocable federated RF.
Then, a suite of homomorphic encryption based secure protocols are designed for federated RF construction, prediction and revocation.
Through theoretical analysis and experiments, we show that the protocols can securely and efficiently implement collaborative training of an RF and ensure that the memories of a revoked participant in the trained RF are securely removed.

\end{abstract}

\keywords{Revocable Federated Learning, Privacy-Preserving, Random Forest}

\maketitle

\section{Introduction} \label{sec_introduction}
\begin{figure}[ht!]
\centering
\includegraphics[scale=0.9]{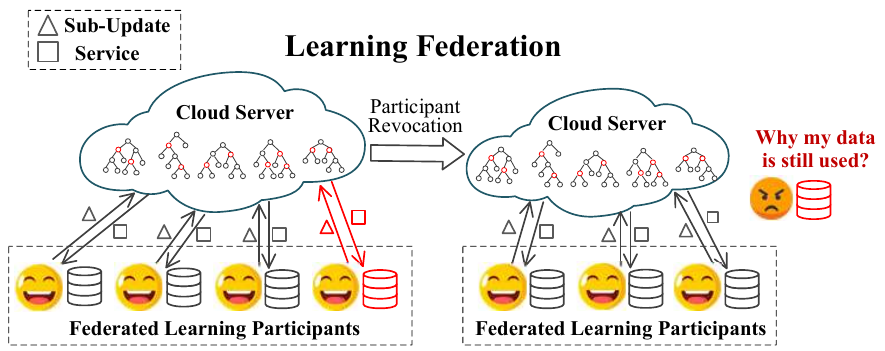}
\caption{A defect of existing FL frameworks}
\label{fig_problem_intro}
\end{figure}

Federated learning (FL) is a novel collaborative learning framework proposed by Google \cite{mcmahan2016communication}.
As shown in Fig.~\ref{fig_problem_intro}, its main idea is to build a machine learning model with the sub-updates derived from distributed datasets, which accelerates model training speed and avoids direct privacy leakage.
Benefited from the artful design, FL attracts increasing interest among scholars and is widely applied to all kinds of application scenarios, such as speech recognition \cite{mcmahan2017learning}, e-health \cite{brisimi2018federated} and especially, mobile networks \cite{bonawitz2017practical}.
Moreover, as the data reconstruction attack towards the sub-updates is discovered \cite{hitaj2017deep}, scholars also put a consistent concern to enhance the privacy-preserving strategy in FL \cite{bonawitz2017practical, bonawitz2019towards}.
Commonly, the core of the privacy-enhancing frameworks is using the cryptographic tools (e.g., homomorphic encryption \cite{cheon2018bootstrapping}) to implement secure aggregation of sub-updates, which can greatly increase the hardness of launching the data reconstruction attack \cite{wang2019beyond}.

However, all the existing frameworks are based on a default assumption that a participant never leaves from the federation or only temporarily loses connection.
This means once a participant is involved in a federation and has ever uploaded a sub-update, its ``trace'' will be permanently remained in the trained model, as shown in Fig.~\ref{fig_problem_intro}.
Recent researchs~\cite{salem2018ml, carlini2019secret, truex2019demystifying} points out that the ``trace'' can leave a chance for the adversary to infer a participant's data, even though the participant has been revoked from the learning federation.
Naturally, such a kind of privacy leakage is unfair and unacceptable for a revoked participant.
Furthermore, in a real-world setting, the defect can leave many potential pitfalls.
A typical example is the cooperation of multiple companies (e.g., hospitals mentioned in \cite{brisimi2018federated}) based on the FL technique.
If one of the companies ends the cooperation with others, the subsequent usage of the information derived from its business data in the learning federation is obliviously illegal.
This kind of dispute can greatly hinder the further development of FL.

Inspired by the above discussion and the basic FL framework, we think that a benchmarking learning federation should satisfy the following security requirements.
\begin{itemize}
    \item \textit{Collaboration Privacy.}
    The original data of a participant cannot be directly or indirectly revealed to others,
    especially in the gradient aggregation process.
    
    \item \textit{Usage Privacy.}
    The machine learning model built by the FL technique is sometimes treated as a publicly-available ``infrastructure'' of the learning federation.
    Therefore, there are two security requirements for usage privacy:
    1) ensuring that no original data are revealed in the usage stage;
    2) protecting the privacy of usage request content.
    
    \item \textit{Revocation Privacy.} 
    The revoked participant has the right to choose whether to leave its contributed data in the learning federation or not.
    If the choice is ``no'', the data of the revoked participant should be neither available for the remaining participants nor remembered by the trained model.
\end{itemize} 

Up to now, most FL frameworks can achieve the collaboration privacy \cite{bonawitz2017practical, mcmahan2017learning, mcmahan2017learning}, and some of them also consider the second goal \cite{liu2019boosting, liu2019lifelong}
Nevertheless, none of them defines or resolves the potential problems brought by participant revocation.
Therefore, we introduce the revocable FL concept in this paper and propose \sysname, a revocable federated random forest (RF) framework to further illustrate the concept.
\sysname achieves federated RF construction and prediction with the homomorphic encryption technique, and meanwhile, supports secure participant revocation.
Besides the high popularity, the reason for choosing RF as our target model is that compared with the other models, like the neural network, the tree structure of RF is more intuitive to state the revocation privacy problem of FL.
Our contributions in this paper are as follows.
\begin{enumerate}
    \item \textbf{Revocable Federated Learning.}
    \sysname extends the practicality of FL in real-world scenarios by introducing the revocation concept.
    To more intuitively illustrate the concept of revocable federate learning, \sysname further defines a revocable and efficient federated RF framework.
    
    \item \textbf{Secure RF Construction.}
    Based on FL, \sysname implements RF construction without direct local data revealing.
    Different from the traditional privacy-preserving RF schemes, the tree nodes of RF in \sysname are from different participants and encrypted with different public keys, which is the basis of realizing the participant revocation security.
    
    \item \textbf{Secure RF Prediction.}
    Based on the homomorphic encryption technique, \sysname ensures that RF prediction can be completed without revealing any information about the prediction request and the RDT nodes, which well meets the security requirements for usage privacy.
    
    \item \textbf{Secure Participant Revocation.}
    Based on a specially designed participant revocation protocol, \sysname implements two levels of revocation.
    For the first-level, we ensure that the data of an ``honest'' revoked participant are securely removed from the learning federation.
    For the second-level, we further ensure that even a revoked participant is corrupted, its data are still unavailable for the adversary.
    
    \item \textbf{Low Performance Loss and High Efficiency.}
    We conduct experiments to prove that \sysname only causes less than 1\% performance loss during RF construction, and costs about $3$ seconds to construct an RDT (faster about 1000 times than existing privacy-preserving RF frameworks).
\end{enumerate}

\noindent
\textbf{Outline.}
In Section~\ref{sec_related_work}, we discuss the related work and background of \sysname.
In Section~\ref{sec_system}, we define the system and security models of \sysname.
In Section~\ref{sec_premilinary}, we describe the cryptographic tools used in \sysname.
In Section~\ref{sec_approach}, the implementation details of \sysname are presented.
Section~\ref{sec_analysis} proves the security of \sysname in a \textit{curious-but-honest} model.
Followed by the comprehensive experiment in Section~\ref{sec_experiments}, the last section concludes this paper.


\section{Related Work and Background} \label{sec_related_work}
In this section, we briefly review the related work and background knowledge of federated learning and random forest (RF).

\subsection{Federated Learning}
Federated learning (FL) is a decentralized machine learning framework that is originally designed to achieve collaborative learning with mobile users~\cite{mcmahan2016communication}.
For FL, one of the biggest advantage is that the attack surface is limited to the device layer, which dramatically reduces the risk of privacy leakage.
Recent work of Bonawitz \textit{et al.} \cite{bonawitz2019towards} claimed that they have successfully applied the FL technique over tens of millions of real-world devices and anticipated billion-level uses in the future.
Moreover, besides the deep neural network (DNN) applied in the original framework~\cite{mcmahan2016communication}, FL is also extended to other machine learning models, such as support vector machine (SVM) \cite{smith2017federated}, long short term memory network (LSTM) \cite{mcmahan2017learning} and extreme gradient boosting forest (XGBoost) \cite{liu2019boosting}.

Furthermore, the original FL is designed without any protection of cryptographic tools~\cite{mcmahan2016communication}.
The later researches~\cite{hitaj2017deep, wang2019beyond} pointed out that the naive design of FL is no longer secure with the existence of dishonest participants.
Thus, some cryptographic tools are added in FL against these attacks~\cite{bonawitz2017practical, mcmahan2017learning, hardy2017private}.
Bonawitz \textit{et al.}~\cite{bonawitz2017practical} utilized the secret sharing technique (SS) to complete the secure aggregation of the secretly shared sub-updates.
However, the most significant problem of SS is that it is vulnerable to the collusion attack ~\cite{yi2015practical}.
Mcmahan \textit{et al.}~\cite{mcmahan2017learning} proposed a federated language recognition model with the differential privacy technique (DP).
For DP, it is a quite difficult task to balance the trade-off between performance loss and efficiency \cite{jayaraman2019evaluating}.
Additionally, Hardy \textit{et al.}~\cite{hardy2017private} designed a private FL framework towards the vertically partitioned data with homomorphic encryption (HE).
Compared with the other two cryptographic tools, HE is thought to have less performance loss and be more robust to the collusion attack~\cite{yang2019federated}.


\subsection{Random Forest}
RF is an ensemble machine learning model that contains multiple random decision trees (RDTs) trained by the bagging method \cite{denil2014narrowing}.
In this paper, we use RF to give a benchmark of revocable FL.
For RF training, the most important operation is the leaf expansion \cite{denil2014narrowing}.
There are two parts to specify a leaf expansion method, which are candidate split recommendation and candidate split quality assessment.
In \sysname, these two parts are completed based on the RF framework proposed in  \cite{denil2014narrowing}, which outperforms the standard RF framework.

As one of the most widely applied machine learning models, how to implement privacy-preserving RF has been a research hotspot.
In 2013, Vaidya \textit{et al.}~\cite{vaidya2013random} proposed a HE based framework that implements both privacy-preserving RF construction and prediction.
Followed by Vaidya's work, Ma \textit{et al.}~\cite{ma2019privacy} designed a high-accurate privacy-preserving RF framework for the outsourced disease predictor.
The traditional HE based frameworks have a common defect that they encrypt the whole database in the RF construction stage, which is inefficient.
Rana \textit{et al.}~\cite{rana2015differentially} used the DP technique to implement a more efficient privacy-preserving RF framework.
Nonetheless, as discussed before, the DP technique usually introduces obvious performance loss.
Moreover, there are also some privacy-preserving RF frameworks \cite{bost2015machine, wu2016privately} which are only designed for prediction and not enough for practical use.

\section{System Model of \sysname}\label{sec_system}
In this section, we give the design motivation of \sysname and define its system model and security model.


\subsection{motivation} \label{sub_motivation}

\begin{figure}[ht!]
\centering
\includegraphics[scale=0.65]{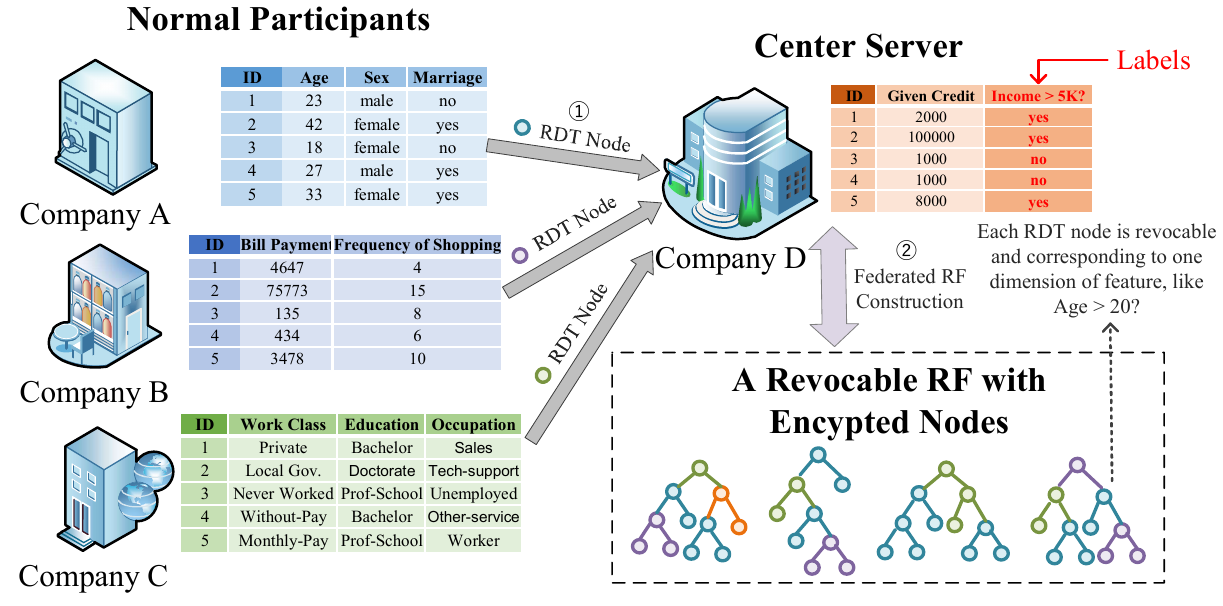}
\caption{A real-world application scenario of FL (predicting whether a customer's monthly income is more than 5K)}
\label{fig_problem_example}
\end{figure}

Our research is mainly motivated by a real-world application scenario of FL, shown in Fig.~\ref{fig_problem_example}, which is a company-level cooperation. 
Different from the user-level cooperation presented in the original FL framework \cite{mcmahan2016communication}, the data shared in the company-level cooperation has considerable commercial values.
Therefore, once a company ends the cooperation with others, it cannot want its data to be still used by other companies.
To more clearly state the problem, RF is obviously the best choice among the massive machine learning models because of its intuitive tree structure.
As shown in Fig.~\ref{fig_problem_example}, the ``memory'' of an RF is directly reflected in the form of RDT node.
To let a participant be able to securely quit from a learning federation, the most intuitive method is to delete the parts (RDT nodes) of the model influenced by the revoked participant and ensure the deleted parts no longer available for the remaining participants.
The first goal is easy to implement.
Nevertheless, to achieve the second goal, we must resort to the cryptographic tool.
Consequently, we introduce a modified Distributed Two Trapdoors Public-Key Cryptosystem (DT-PKC, defined in Section~\ref{sec_premilinary}) \cite{bresson2003simple} in federated RL and propose a benchmark of federated RF framework, \sysname. 
Since DT-PKC supports homomorphic encryption across different domains, \sysname ensures that the RDT node that is encrypted with its provider's key can still be computable for RF prediction, and also, easily removed from an RF.
The details of \sysname are presented in Section~\ref{sec_approach}.

\subsection{System Model}
\begin{figure}[ht!]
\centering
\includegraphics[scale=0.78]{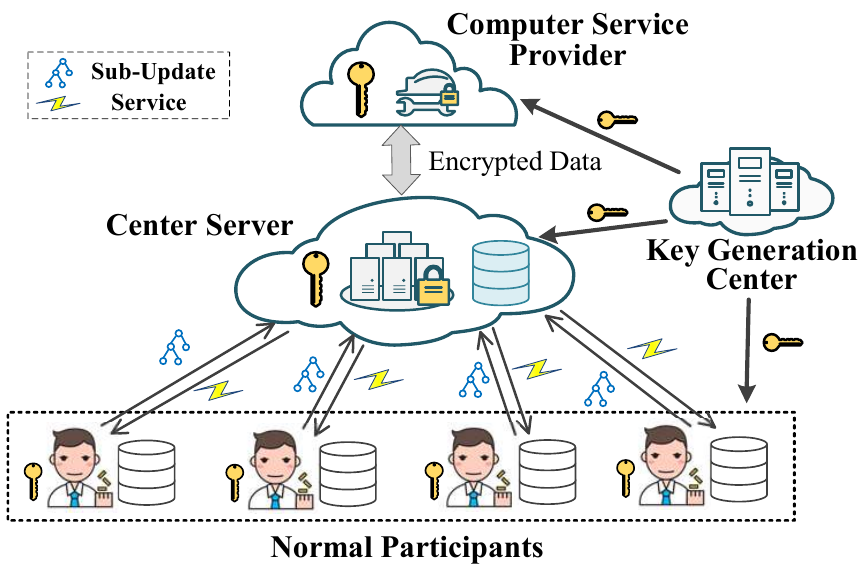}
\caption{The system model of \sysname}
\label{fig_system_model}
\end{figure}

As shown in Fig.~\ref{fig_system_model}, \sysname comprises four kinds of entities, namely a center server (CS), a set of normal participants (UD), a computation service provider (CC) and a key generation center (KGC).
The data in \sysname is vertically partitioned, as illustrated in Fig.~\ref{fig_problem_example}

\textbf{Center Server.}
\entity{CS} is usually an initiator of a learning federation of \sysname.
It takes on most of the computation tasks in \sysname and manages the usage of the trained RF model.
Specially, \entity{CS} is also a data provider who has the ground truths, i.e., the labels for classification or the prediction target for regression.

\textbf{Normal Participant.}
\sysname involves more than one normal participants, \entity{UD} $= \{u_1, u_2, ...,\}$.
Each $u_i\in$ \entity{UD} has one or more dimensions of data used for RF construction.

\textbf{Computation Service Provider.}
\entity{CC} is responsible for assisting \entity{CS} to complete the complex computations of HE.

\textbf{Key Generation Center.}
\entity{KGC} is only tasked with key generation and distribution.

\subsection{Adversary Model}\label{sub_security_model}
In \sysname, \entity{KGC} is a trusted party.
A trusted party always honestly completes its task and does not collude with anyone else.
\entity{CS}, \entity{UD} and \entity{CC} are \textit{curious-but-honest}, which means they follow the promised steps of our protocols but also want to benefit themselves by learning other parties' data.
Therefore, we introduce a \textit{curious-but-honest} adversary $\mathcal{A}$.
$\mathcal{A}$ is restricted from compromising both \entity{CS} and \entity{CC} but can corrupt any subset of \entity{UD}.
For the corrupted participants, $\mathcal{A}$ obtains their local data and private keys.
In the RF construction and predication stages, the goal of $\mathcal{A}$ is using the obtained knowledge to derive the private data of ``honest'' participants.
The private data include the original feature data and the candidate splits, because .
Specially, in the participant revocation stage, we define two levels of revocation strategies.
For the first-level revocation, $\mathcal{A}$ is restricted from colluding with the revoked participant.
We call the security implemented by the first-level revocation as ``forward'' security.
For the second-level revocation, $\mathcal{A}$ is released from the restriction.
Correspondingly, the security implemented by the second-level revocation is called ``backward'' security.
In the two levels of revocation, the goals of $\mathcal{A}$ are identical, which are deriving the private data of `honest'' participants and operating the old RF without the revoked participant.
Different from the former two stages, $\mathcal{A}$ has an additional goal in this stage.
This is because \sysname has to rebuild the RF after a participant is revoked.
If the old RF is available, the revocation naturally becomes meaningless.

\section{Cryptographic Tools} \label{sec_premilinary}
The security of \sysname is mainly provided by HE.
\sysname introduces the DT-PKC for operating homomorphic encryption.
Table~\ref{table_notation} summarizes the frequently-used notations of HE.
Three types of keys are generated in a DT-PKC, namely public key, weak private key and strong private key.
The public key is used for encryption.
The weak key and the strong private key are used for partial decryption and full decryption in \sysname, respectively.
Their generation process is as follows.
\renewcommand\arraystretch{1}
\begin{table}[!htbp]
\centering
\caption{Notation Table}
\begin{tabular}{l l}

\specialrule{.1em}{.1em}{.1em}
\textbf{Notations}       & \textbf{Descriptions} \\

\hline



$pk_u, sk_u$             & A pair of public-private keys for HE.\\

$\lambda_1, \lambda_2$   & Randomly split strong private keys for HE. \\

$[\![m]\!]_{pk_u}$       & A ciphertext encrypted with $pk_u$ using HE.\\

$|\cdot|$                & The size of an arbitrary set.\\

$||\cdot||$              & The data length of an arbitrary variable.\\

$\kappa$                 & The security parameter.\\

$p$,$q$                  & $p$ and $q$ are two big primes, $||p|| = ||q|| = \kappa$.\\

$N$                      & $N$ is a big integer satisfying $N = pq$.\\

$\mathbb{Z}_N$           & An integer field of $N$.\\

\specialrule{.1em}{.1em}{.1em}
\end{tabular}
\label{table_notation}
\end{table}

Given a security parameter $k$ and two arbitrary large prime numbers $p$, $q$, we first derive another two numbers $p' = (p - 1)/2$ and $q' = (q - 1)/2$, where the data lengths of $p$ and $q$ are both $k$.
Then, we compute a generator $g = -a^{2N}$ of order $(p - 1)(q - 1)/2$, where $a\in \mathbb{Z}_{N^2}$ is a random number.
Finally, a weak private key $sk$ is randomly selected from $[1, N/4]$ and its corresponding public key is computed by $pk = (N, g, h = g^{sk})$.
The strong private key is $\lambda = lcm(p - 1, q - 1)$, where $lcm(\cdot)$ is a function to compute the lowest common multiple.
In \sysname, $\lambda$ is randomly split into $\lambda = \lambda_1 + \lambda_2$.
$\lambda_1$ and $\lambda_2$ are distributed to \entity{CS} and \entity{CC}, respectively.

There are five DT-PKC based functions \cite{liu2016efficient} involved in \sysname, including encryption (HoEnc), re-encryption (HReEnc), ciphertext refresh (HEncRef), partial decryption (ParHDec1 and ParHDec2) and comparison (HoLT).
Their detailed implementations are presented in Appendix~\ref{appendix_functions}.

\begin{enumerate}
    \item \textbf{Encryption.}
    Given a plaintext message $m$ and a public key $pk_{u_1}$, \func{HoEnc}$(pk_{u_1}, m)$ outputs a ciphertext $[\![m]\!]_{pk_{u_1}}$.
    
    \item \textbf{Re-Encryption.}
    Given a ciphertext $[\![m]\!]_{pk_{u_1}}$, $u_2$ can use $pk_{u_2}$ to compute \func{HReEnc}$(pk_{u_2}, [\![m]\!]_{pk_{u_1}})$ and output a re-encrypted ciphertext $[\![m]\!]_{pk_{\Sigma}}$, where $pk_{\Sigma} = pk_{u_1} + pk_{u_2}$.
    
    \item \textbf{Ciphertext Refresh.}
    Given a ciphertext $[\![m]\!]_{pk_{u_1}}$, \func{HEncRef}$(r,$ $ [\![m]\!]_{pk_{u_1}})$ refreshes the ciphertext without changing the plaintext, where $r$ is randomly chosen from $\mathbb{Z}_N$.
    
    \item \textbf{Partial Decryption.} 
    Partial decryption contains two steps.
    Given a ciphertext $[\![m]\!]_{pk_{\Sigma}}$, \func{ParHDec1} $(sk_{u_1}, [\![m]\!]_{pk_{\Sigma}})$ outputs a partially decrypted result $[\![m]\!]_{pk_{u_2}}$;
    \func{ParHDec2}$(sk_{u_2}, [\![m]\!]_{pk_{u_2}})$ outputs the plaintext message $m$.
    
    \item \textbf{Comparison.}
    Given two ciphertexts $[\![m_1]\!]_{pk_{u_1}}$ and $[\![m_2]\!]_{pk_{u_2}}$, $u_3$ uses \func{HoLT}$([\![m_1]\!]_{pk_{u_1}}, [\![m_2]\!]_{pk_{u_2}})$ to output an encrypted result $[\![l]\!]_{pk_{\Sigma}}$, where $pk_{\Sigma}$ can be $pk_{u_3} + pk_{u_2}$ or $pk_{u_3} + pk_{u_1}$.
    If $m_1$ is less than $m_2$, $l$ is $1$; otherwise, $l$ is $0$.
\end{enumerate}

In these functions, the plaintext is within $\mathbb{Z}_N$ and the ciphertext is within $\mathbb{Z}_{N^2}$.
The areas $[0, R_1)$ and $(N - R_1, N]$ are used to represent the positive and negative numbers, respectively, where $||R_1|| < ||N||/4$.
Specially, the numbers need to be encrypted can be not integers.
To resolve the problem, we use the fixed-point format to represent these numbers.
For brevity, the detailed discussion about the data format is given in Appendix~\ref{appendix_fixed_point}.


\section{\sysname Framework}\label{sec_approach}
In this section, we first overview the workflow of \sysname, and then, present its implementation details.
Notably, for easy understanding of readers, we suppose that one participant only has one dimension of feature data in this section.
\begin{algorithm*}[ht!]
  \caption{Federated Leaf Expansion (\protocol{FLE-Expan})}
  \label{pro_fle_expan}
  \begin{algorithmic}[1]
    \Require
    A sample selection vector $\vec{\mu}$;
    a randomly selected subset $\mathcal{F}'$ of the feature set $\mathcal{F}$;
    the task type $task \in \{0, 1\}$;
    current tree depth $d_c$.
    \Ensure
    A random decision tree $T$.

    \If{$d_c > d_{max}$ or $\mathcal{F}'$ is empty}
        \State \Return A random decision tree $T$.
    \EndIf

    \State \entity{CS} generates a feature selection vector $\vec{v} = (v_1, v_2, ..., v_{|\mathcal{F}|})$. If feature $i$ belongs to both $\mathcal{F}'$ and $\mathcal{F}$, $v_i = 1$; otherwise, $v_i = 0$.

    \State \entity{CS} distributes the sample selection vector $\vec{\mu}$ and the feature selection vector $\vec{v}$ to each $u\in$ \entity{UD}.

    \For{$u_{\tau}\in$ \entity{UD} and $v_{\tau}\in \vec{v}$ is $1$}


            \State Choose the selected samples $D_{\tau}'$ based on $\vec{\mu}$.

            \State Uniformly and randomly select $\varrho$ samples from the selected samples, i.e., $D_{\tau}'' = \{x_{\tau, 1}, x_{\tau, 2}, ..., x_{\tau, \varrho}\}$ and $D_{\tau}''\subset D_{\tau}'$.

            \State Compute $x_{min} = \mathop{\arg\min}_{x_{\tau, k}\in D_{\tau}''}x_{\tau, k}$ and $x_{max} = \mathop{\arg\max}_{x_{\tau, k}\in D_{\tau}''}x_{\tau, k}$.

            \State Averagely pick $\varsigma$ candidate splits $\mathcal{S} = \{s_1, s_2, ..., s_{\varsigma}\}$ in the interval $[x_{min}, x_{max}]$.

            \State Towards each candidate split $s_{\imath}\in \mathcal{S}$, initialize a split vector $\vec{w}_{\imath} = (w_{\imath, 1}, w_{\imath, 2}, ..., w_{\imath, |D_{\tau}|})$.
            For the unselected sample $x_{\tau, \jmath}\notin D_{\tau}''$, $w_{\imath, \jmath}$ is set to $0$.
            For $x_{\tau, \jmath} \leq s_{\imath}$, set $w_{\imath, \jmath} = -1$, otherwise, $w_{\imath, \jmath} = 1$.
            In this way, $u_{\tau}$ can generate a selection set $\mathcal{W}_{\tau} = \{\vec{w}_1, \vec{w}_1, ..., \vec{w}_{\varsigma}\}$.


    \EndFor


    \State For each each participant $u_{\tau}\in$ \entity{UD}, \entity{CS} collects $\mathcal{W}_{\tau}$ and constructs $\mathcal{W} = \{\mathcal{W}_1, \mathcal{W}_2, ..., \mathcal{W}_{|UD|},\}$.

    \State \entity{CS}
    removes the feature corresponding to $s_p$ from $\mathcal{F}'$,
    adds $(split, \vec{u}, \vec{w}_0, d_c, u_0, \sigma_0, \mathcal{F}') = $\protocol{POS-Find}$(\mathcal{W}, task)$ into current tree,
    updates $d_c = d_c + 1$,
    invokes \protocol{FLE-Expan}$($\func{Sign}$(\vec{w}_0), \mathcal{F}', task, d_c)$ and \protocol{FLE-Expan}$($\func{Sign}$(-\vec{w}_0), \mathcal{F}', task, d_c)$ to generate two child nodes.

  \end{algorithmic}
\end{algorithm*}

\subsection{\sysname Overview}\label{sub_overview}
\begin{figure}[ht!]
\centering
\includegraphics[scale=0.8]{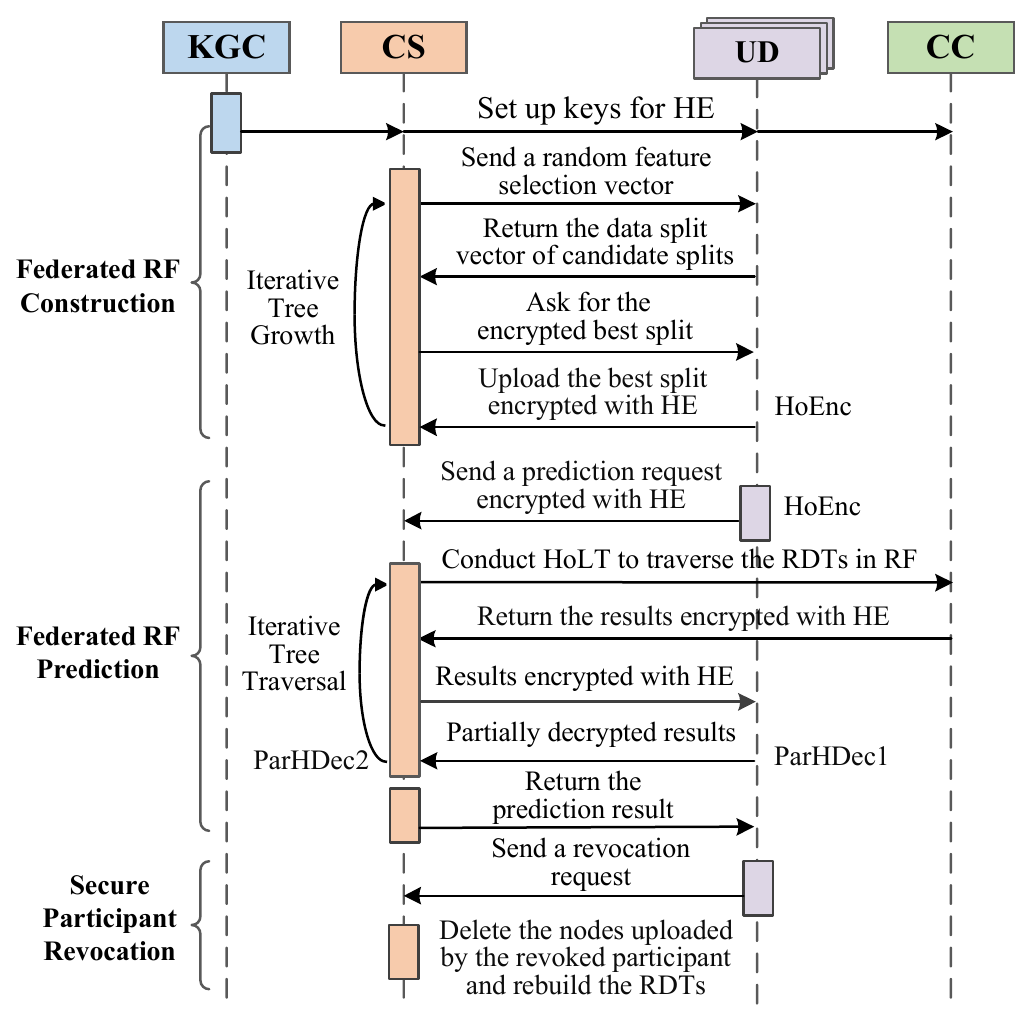}
\caption{High-level overview of \sysname}
\label{fig_workflow}
\end{figure}

As shown in Fig.~\ref{fig_workflow}, \sysname contains three stages, namely secure RF construction, secure RF prediction and secure participant revocation.
A brief overview of these stages are given below.

\noindent
\textbf{Secure RF Construction.}
\sysname constructs an RF in two steps.

1. \textit{Key Setup.}
\entity{KGC} generates the HE keys and distributes them to all participants.

2. \textit{Federated Tree Growth.}
To avoid the private data of participants revealed to others, the tree growth in \sysname is implemented by iteratively invoking a specially designed leaf expansion protocol.
In the protocol, \entity{CS} first randomly chooses a subset of all features.
Then, the normal participants that own the data of chosen features recommend candidate splits in a random range.
The candidate splits are not directly sent to \entity{CS} but sent in split vector format (equal to the gradient in DNN~\cite{mcmahan2016communication}) to avoid privacy leakage.
Finally, \entity{CS} asks the participant who provides the split vector with the highest quality to upload the encrypted best split.
The encrypted best split is stored as a new RDT node in the fixed-pointed format (defined in Appendix~\ref{appendix_fixed_point}).

\vspace{0.1cm}
\noindent
\textbf{Secure RF Prediction.}
Secure RF prediction is used to securely process a prediction request.
The request is not vertically partitioned and contains all dimensions of features.
Such a request type is commonly discussed in the privacy-preserving RF frameworks \cite{vaidya2013random, bost2015machine, wu2016privately}.
The processing procedures of a prediction request are given below.
First, the requester sends the encrypted prediction request to \entity{CS}.
Then, \entity{CS} iteratively traverses the RF with a secure RF prediction protocol.
Finally, \entity{CS} computes the prediction result according to the task type and returns it to the requester.

\vspace{0.1cm}
\noindent
\textbf{Secure Participant Revocation.}
Secure participant revocation guarantees that the data of revoked participants are removed from the RF and no longer available for remaining participants.
When a normal participant wants to quit a learning federation, \entity{CS} first traverses the RDTs in the trained RF and destroys the splits provided by the participant.
Then, the protocol used for RF construction is invoked to rebuild the destroyed RDT nodes.
In most cases, these steps are enough to provide ``forward'' security of participant revocation.
This is because the RDT nodes in \sysname are always kept in encrypted format and can only be used with the existence of their providers.
If we suppose the revoked participant is ``honest'', the above steps have been able to ensure the revoked data to be unavailable for the subsequent use of the old RF.
Nonetheless, consider the situation where the adversary corrupts the participant after it is revoked.
The above revocation is no longer secure from the ``backward'' perspective.
By colluding with \entity{CS} and the revoked participant, the adversary can operate the old RF without being noticed by the other ``honest'' participants.
Therefore, we further provide a second-level revocation to ensure ``backward'' security.
Compared to the first-level revocation, extra computations are involved in the second-level revocation to refresh the revoked splits with random values.
By this way, we ensure that the revoked splits are no longer available even for its provider.





\subsection{Secure RF Construction}\label{sub_rf_con}
\sysname mainly uses two secure protocols for RF construction, namely the federated leaf expansion protocol (FLE-Expan) and the federated optimal split finding protocol (FOS-Find).
The two protocols allow the participants of \sysname to collaboratively construct an RF without directly uploading their local data.
The following is the implementation details of secure RF construction.







\textbf{Key Setup.}
Before constructing the RDTs of RF, \entity{KGC} first initializes the cryptographic keys.
The key distribution can be completed offline or by secure channels.

\textbf{Federated Tree Growth.}
In \sysname, \entity{CS} grows a RDT by iteratively invoking \protocol{FLE-Expan}.
Every time \protocol{FLE-Expan} is completed, a RDT node is expanded in current RDT.
To grow a RDT with depth $d_{max}$, \protocol{FLE-Expan} has to be operated at most $\sum_{i = 0}^{d_{max}} 2^i$ times.
Two parts specify the RF construction process.
\begin{figure}[ht!]
\centering
\includegraphics[scale=0.65]{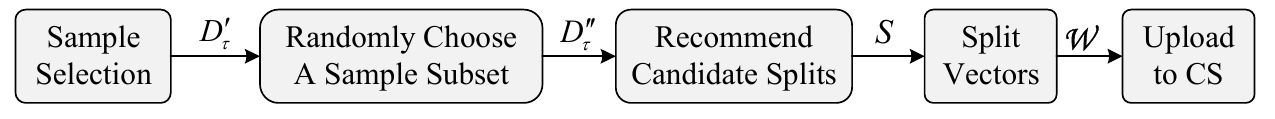}
\caption{The workflow of leaf expansion.}
\label{fig_leaf_expan}
\end{figure}

One is the recommendation of candidate splits (Protocol~\ref{pro_fle_expan}, line 4-18).
Its workflow is as illustrated in Fig.~\ref{fig_leaf_expan}.
\entity{CS} first randomly selects a feature subset $\mathcal{F}'\subset \mathcal{F}$ and builds a feature selection vector $\vec{v}$ corresponding to the selected features.
The size of $\mathcal{F}'$ is usually recommended to be $\sqrt{|\mathcal{F}|}$~\cite{biau2016random}.
Then, the selection vector is distributed to all participants.
Each participant checks whether its feature is selected.
If selected, the participant $u_{\tau}$ confirms the involved samples of current node, $D_{\tau}'$, through a sample selection vector $\vec{\mu}$.
Each dimension of $\vec{\mu}$ corresponds to a training sample and is initially set to $1$.
Next, \entity{CS} randomly picks out a subset of the involved samples, $D_{\tau}''$, and determines its minimum value $x_{min}$ and maximum value $x_{max}$.
The candidate splits $\mathcal{S}$ are averagely chosen between $x_{min}$ and $x_{max}$.
Corresponding to each candidate split, $u_{\tau}$ computes a 0-1 split vector, $\vec{w_i}$.
Finally, the set of split vectors $\mathcal{W} = \{\vec{w_1}, \vec{w_2}, ...\}$ is sent it to \entity{CS} for split quality assessment.
\begin{algorithm}[ht!]
  \caption{Federated Optimal Split Finding (\protocol{FOS-Find})}
  \label{pro_fos_find}
  \begin{algorithmic}[1]
    \Require
    The split vector set $\mathcal{W}$;
    the task type $task \in \{0, 1\}$;
    \Ensure
    An optimal RDT node.

    \State For each $\vec{w}\in \mathcal{W}$, \entity{CS} splits its local data set by computing $D_1 = $\func{Sign}$(\vec{w})\times D$ and $D_2 = $\func{Sign}$(-\vec{w})\times D$.

    \If{$task$ is $0$} \# regression task
        \State $\vec{w}_0 = \mathop{\arg\min}_{\vec{w}\in \mathcal{W}} E(D_1, D_2, \vec{w})$.
    \Else \xspace \# classification task
        \State $\vec{w}_0 = \mathop{\arg\min}_{\vec{w}\in \mathcal{W}} G(D_1, D_2, \vec{w})$.
    \EndIf

    \State \entity{CS} asks the participant $u_0$ that provides $\vec{w}_0$ to upload the corresponding split $s_0$.

    \State $u_0$ encrypts $s_p = [\![s_0]\!]_{pk_0} = $\func{HoEnc}$(pk_0, s_0)$, and sends $[\![s_0]\!]_{pk_0}$ to \entity{CS}.

    \State \Return the optimal split $s_p$.
  \end{algorithmic}
\end{algorithm}

The other is the assessment of candidate split quality (Protocol~\ref{pro_fos_find}).
Based on $\mathcal{W}$, \entity{CS} completes the assessment by invoking \protocol{FOS-Find}.
During the assessment, we introduce the following sign function.
\begin{equation}\label{eq_sign_function}
    sign(x) =
    \begin{cases}
        1, & \mbox{if } x > 0 \\
        0, & \mbox{if } x \leq 0
    \end{cases}.
\end{equation}
\entity{CS} inputs each element of $\vec{w}\in \mathcal{W}$ into \func{Sign}$(\cdot)$.
If the output is $1$, the sample corresponding to the element is added in the left child, otherwise, it is added in the right child.
Towards different tasks, the quality of a candidate split is assessed in different ways.
For regression, \entity{CS} computes the mean squared errors (MSE) by $E(D_1, D_2, \vec{w})$.
For classification, \entity{CS} computes the Gini coefficients by $G(D_1, D_2, \vec{w})$.
The computation methods of the two functions are given in Appendix~\ref{appendix_quality_assessment}.
Among all candidate splits, the one with the lowest MSE or Gini coefficient is chosen as the optimal split of the current node.
And \entity{CS} asks the participant that provides the optimal split to upload it in encrypted format.
\protocol{FLE-Expan} is iteratively invoked until reaching the maximum tree depth $d_{max}$.


\subsection{Secure RF Prediction} \label{sub_rf_pre}
In \sysname, a participant can use the constructed RF by invoking the federated prediction protocol (FRF-Predict and FT-Predict).

The requester $u_0$ encrypts his request with his public key and sends it to \entity{CS}.
As received the encrypted request, \entity{CS} repeatedly operates \protocol{FT-Predict} to get the prediction result $r_i$ of each RDT.
The procedure of \protocol{FT-Predict} is as follows.
At the beginning of the root node of tree $T_i$, \entity{CS} extracts the encrypted split $[\![s_{\tau}]\!]_{u_{\tau}}$ and its corresponding provider $u_0$.
By invoking \func{HoLT}, \entity{CS} can obtain the comparison result $l$ between the feature value in the request and the split.
The comparison result is encrypted with both the public keys of $u_{\tau}$ and \entity{CS}.
Therefore, both $u_{\tau}$ and \entity{CS} are necessary to obtain the plaintext comparison result (Protocol~\ref{pro_ft_predict}, line 8-9).
Finally, if the decrypted comparison result $l$ is $1$, \entity{CS} enters the left child node; otherwise, enters the right child node.
Notably, in this process, the 0-1 output is revealed to \entity{CS}.
The revealing of 0-1 output does not influence the security \sysname, because it tells nothing but the relation between RDT node and the request are both encrypted.
According to \cite{liu2016efficient}, the information is not enough to derive the plaintext data.
The above operations are repeated until reaching a leaf node. 
As all RDTs are traversed, \entity{CS} computes the final output $O$ and returns it to the requester.
The final output has two types.
For regression, $O$ is a mean value of each RDT output.
For classification, $O$ is the category with the most votes among all RDTs.

Sometimes, \sysname also has to process the vertically partitioned data for model testing during RF construction.
The processing of the prediction for testing is similar to the above steps except that the comparison operations during the RF traversal can be locally completed by each participant. 
Therefore, we do not discuss this type of prediction in details and only present its implementation in Appendix~\ref{appendix_sec_prediction} ( Protocol~\ref{pro_frf_testing} and Protocol~\ref{pro_ft_testing}).


\subsection{Secure Participant Revocation} \label{sub_rf_revo}
\sysname implements participant revocation with the secure participant revocation protocol (FRF-Revoc, Protocol~\ref{pro_frf_revoc}).
\protocol{FRF-Revoc} provides two levels of participant revocation.
For the first-level revocation, we guarantee that the data of the revoked participant are no longer available by the remaining participants, and the RF is securely rebuilt after the revocation.
If the revoked participant is ``honest'', such level of revocation is secure enough.
However, if both \entity{CS} and the revoked participant are corrupted, the revocation becomes insecure.
According to our assumption, the adversary can obtain and copy the data stored in a corrupted \entity{CS}, which means that simply asking \entity{CS} to destroy the data of the revoked participant is meaningless.
With the copied data and the corrupted private key of the revoked participant, the adversary can totally operate the old RF model with remaining ``honest'' participants.
This is because the ``honest'' participant only partially decrypts the messages received in the RF prediction stage, and cannot identify whether the running RF is old or not.
Therefore, we further propose the second-level revocation to ensure that the data about the revoked participants are not available even by their provider to avoid the ``backward'' attack.
The revocation process is as shown in Fig.~\ref{fig_revocation} and given below.
\begin{algorithm*}[ht!]
  \caption{Secure Participant Revocation (\protocol{FRF-Revoc})}
  \label{pro_frf_revoc}
  \begin{algorithmic}[1]
    \Require
    The revoked participant $u_r$;
    the original random forest $\mathcal{T} = \{T_1, T_2, ...\}$

    \Ensure
    A rebuilt random forest $\mathcal{T}'$.

    \State $u_r$ sends a revocation request $req_r$ to \entity{CS}.

    \State \entity{CS} computes \func{Verf}$(k_{ver, u_r}, \sigma_r)$.
    If the result is not $1$, reject the request; otherwise, do the following steps.

    \For{$T_i\in \mathcal{T}$}
        \State For each $Node_j \in T_i$ that is not rebuilt, do the following computations.

        \State \entity{CS} extracts the provider $u_j$ of $Node_j$.

        \If{$u_j$ is the revoked participant}
            \State \entity{CS} traverses the child nodes of $Node_j$ until reaching the leaf node.
            All these child nodes and $Node_j$ are destroyed from $T_i$.

            \State To rebuild the destroyed nodes, \entity{CS} invokes \protocol{FLE-Expan}$(\vec{u}, \mathcal{F}', task, d_c)$.
            The inputs of \protocol{FLE-Expan} are previously stored in $Node_j$ (Protocol~\ref{pro_fle_expan}, line 14).

            \State \textbf{For second-level revocation, \entity{CS} still has to do the following three additional steps:}

            \State \textcolor{red}{\entity{CS} sends the revocation request $req_r$ and the destroyed splits of the revoked the nodes to \entity{CC}.}

            \State \textcolor{red}{\entity{CC} generates a random value $r_d$ for each destroyed node $Node_d$ and compute \func{HEncRef}$(r_d, s_d)$, where $s_d$ is the split of $Node_d$.
            All the refreshed nodes are returned to \entity{CS}.}

            \State \textcolor{red}{\entity{CS} also generates a random value $r_d'$ for each destroyed node $Node_d$ and computes \func{HEncRef}$(r_d', s_d)$.}
        \EndIf
    \EndFor
    
    \State \entity{CS} forwards the revocation request to \entity{KGC}, and \entity{KGC} revokes the keys of $u_r$.
    
    \State \entity{CS} publishes the message that $u_r$ is removed from the learning federation to every other participant.
  \end{algorithmic}
\end{algorithm*}

\begin{figure}[ht!]
\centering
\includegraphics[scale=0.9]{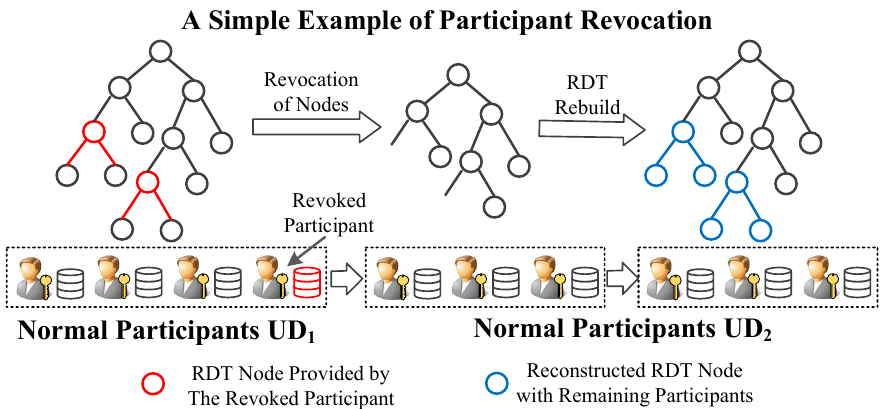}
\caption{A simple example of participant revocation with one RDT}
\label{fig_revocation}
\end{figure}

The revoked participant sends a revocation request to \entity{CS}.
As long as the request is received, \entity{CS} first checks whether the signature is valid.
If the revocation request is valid, \entity{CS} iteratively traverse all RDTs in current RF.
During the traversal, the RDT node provided by $u_r$ and all its child nodes are removed from the RF.
Then, the removed RDT nodes are rebuilt by invoking \protocol{FLE-Expan}.
$u_r$ does not participate in the rebuilding process.
In worst condition (the revoked node is the root node), \entity{CS} has to reconstruct a whole RDT.
Theoretically, the probability of the worst condition is only $\frac{1}{|UD|}$.
Hence, in most cases, the remaining participants can rebuild the federation at a few extra costs.
Moreover, since the RDTs in an RF are isolated from each other, the rebuilding of one RDT does not influence the function of other RDTs.
If we do not consider the ``backward'' security, \entity{CS} publishes the revocation information as soon as the above iterative steps are completed.
Considering the ``backward'' security, we still have to do the following computations to implement the second-level revocation.
While destroying the revoked RDT nodes, \entity{CS} sends the revoked splits to \entity{CC}.
\entity{CC} refreshes the ciphertexts of the splits by computing \func{HEncRefresh}.
The refreshed splits are returned to \entity{CS} and refreshed again.
Here, the revoked splits are refreshed for two times. 
By this way, the splits are encrypted with a public key of its provider and two random keys generated by \entity{CS} or \entity{CC}.
Since \entity{CS} or \entity{CC} cannot be simultaneously corrupted, we guarantee that the refreshed splits are no longer available for the adversary.

Another considerable problem for participant revocation is that the missing of dimensions of data may reduce the effectiveness of the trained model.
A commonly accepted idea is that most features in a dataset are redundant and there is still no way to perfectly eliminate the redundancy \cite{wang2018machine}.
Thus, it is reasonable to think that the lack of a small number of dimensions cannot obliviously the usability of a dataset.
The experiment results in Section~\ref{sec_experiments} further prove the correctness of our thinking (Table~\ref{table_classification_revocation} and Table~\ref{table_regression_revocation}).
Consequently, the revocation of only a few participants does not influence the effectiveness of \sysname in most cases.

\subsection{Further Discussion}\label{sub_further_dis}
Three important security and application problems of \sysname are further discussed below.

\noindent
\textbf{Selection Vectors}
We say that the 0-1 vectors in the RF construction stage do not reveal any information about the participant's private data.
For the feature selection vector $\vec{v}$, it is randomly generated to select a subset of features.
Therefore, it does not relate to any participant's private data.
Then, the sample selection vector $\vec{\mu}$ reflects the sample partition result with the newly obtained best split.
The best split is related to a single dimension of data and the data is only known by its provider.
Thus, $\vec{\mu}$ reveals nothing but the fact that the provider has several secret values and some of them are less than others, which is not enough to derive the original data.
The same explanation can also be used to state that the split vector $\vec{w}_{\imath}\in \mathcal{W}_{\tau}$ does not reveal any private data of $u_{\tau}$.

\noindent
\textbf{Dimension Extension.}
The above implementation is based on an ideal condition where each $u\in$ \entity{UD} owns one dimension of feature data.
However, in most cases, a participant usually has multiple dimensions of data.
To adapt to this condition, we only have to make a few simple modifications to Protocol~\ref{pro_fle_expan}.
First, each $u\in$ \entity{UD} checks more than one dimension of the feature selection vector (Protocol~\ref{pro_fle_expan}, line 7).
Then, for all selected features, $u$ completes the computations of candidate split recommendation (Protocol~\ref{pro_fle_expan}, line 8~13).
Thus, \sysname can handle the multi-dimensions condition and achieve the same performance as before.

\noindent
\textbf{Model Extension.}
In \sysname, we choose RF as a benchmark.
Indeed, other federated learning models, like CNN and DNN, also have the same requirement.
The difference is that for RF, the sub-updates of model training is the candidate splits, while for neural network, the sub-updates are gradients.
The core idea of extending our revocable federated learning concept is derived from the parameter update principle of CNN and DNN, i.e., $\omega_{K} = \omega_0 - \eta \sum_{i = 1}^{K} \sum_{j = 1}^{U_i} \frac{n_j}{n_b} g_j$,
where $\omega_K$ is a parameter after $K$ iterations, 
$\eta$ is the learning rate, 
$U_i$ is the involved participants for $i_th$ iteration, 
$n_b$ is the batch size, 
$g_j$ the average gradient of $n_j$ training samples owned by participant $j$.
To implement revocable federated learning, we just have to let each participant encrypt $g_j$ and use \func{HoAdd} to complete the summing operation.
\func{HoAdd} is a secure HE addition algorithm across different domains, given in Appendix~\ref{appendix_functions}.
Nevertheless, compared with the intuitive tree structure of RF, the structure of the neural network is too abstract.
Therefore, we think RF is more ideal to illustrate our revocable federated learning concept.




\section{Security Analysis}\label{sec_analysis}
In this section, we prove that \sysname is secure under the \textit{curious-but-honest} model.

\subsection{Security of Cryptographic Tools}
To prove \sysname security, we first have to state the security of the utilized cryptographic tools.

\subsubsection{Key Security}
The DT-PKC has two types of trapdoors, weak private key and strong private key.
The weak private key is securely stored by each participant but the strong private key is randomly split and distributed to \entity{CS} and \entity{CC}, respectively.
The security of strong private key split is based on the information-theoretic secure secret sharing framework proposed by Shamir \cite{shamir1979share}.
In \sysname, the key split satisfies $\lambda_1 + \lambda_2 \equiv 0 \mod \lambda$ and $\lambda_1 + \lambda_2 \equiv 1 \mod N^2$.
$\lambda_1$ and $\lambda_2$ are two random shares of the strong private key $\lambda$.
According to the (2, 2)-Shamir secret sharing framework \cite{shamir1979share}, any less than two shares cannot recover the shared value.
Therefore, no matter \entity{CS} or \entity{CC} is compromised by the active adversary (other participants have no knowledge about $\lambda$), $\lambda$ cannot be revealed, which is the following theorem~\cite{liu2016efficient}.
\begin{lemma}\label{thm_key_security}
    The strong private key split described in Section~\ref{sec_premilinary} is derived to be secure from the (2, 2)-Shamir secret sharing under the \textit{honest-but-curious} model.
\end{lemma}


\subsubsection{DT-PKC Security}
DT-PKC has been proved to be semantically secure in the standard model, which is based on the hardness of DDH assumption over $\mathbb{Z}_{N^2}$ \cite{bresson2003simple}.
For brevity, we only give the following lemmas and omit its proof.
\begin{lemma}\label{lem_enc_security}
    The DT-PKC is semantically secure based on the assumed intractability of the DDH assumption over $\mathbb{Z}_{N^2}$, which also derives that the DT-PKC based functions \func{HoEnc}, \func{HReEnc}, \func{HEncRef}, \func{ParHDec1}, \func{ParHDec2} and \func{HoLT} are secure.
\end{lemma}

\subsection{Security of \sysname}
We define the security of our protocols based on a universal composition framework (UC) \cite{mohassel2017secureml}.
According to UC, we assume all participants honestly execute a protocol and there is an \textit{environment} machine called $Env$.
For \textit{honest} participants, their inputs are chosen from $Env$ and their outputs are returned to $Env$.
Without loss of generality, the \textit{curious-but-honest} adversary $\mathcal{A}$ can also interact with $Env$.
Nevertheless, $\mathcal{A}$ only simply forwards all received protocols messages and acts as instructed by $Env$.
For a real interaction of an arbitrary protocol $\pi$, we let $Real(\pi, \mathcal{A}, Env)$ to represent the view of $\mathcal{A}$.
Similarly, $Ideal(\pi, \xi, Env)$ is used to represent the ideal view of $\mathcal{A}$ when we let $Env$ interact with a simulator $\xi$ and \textit{honest} participants.
Based on the above assumptions, a formal definition of protocol security is derived as follows \cite{mohassel2017secureml}.
\begin{definition}\label{def_security}
    A protocol $\pi$ of \sysname is secure in the \textit{curious-but-honest} model if there exist simulators $\xi = \{\xi_{CS}, \xi_{CC}, \xi_{UD}\}$ that can simulate $Ideal(\pi, \xi, Env)$ which is computationally indistinguishable from the real view $Real(\pi, \mathcal{A}, Env)$ of honest-but-curious adversaries $\mathcal{A} = \{\mathcal{A}_{CS}, \mathcal{A}_{CC}, \mathcal{A}_{UD}\}$.
\end{definition}
Definition~\ref{def_security} is the basis of our security proofs of \sysname.
According to the definition, we prove that an adversary cannot obtain more knowledge from received protocol messages than a suite of meaningless random values.
In addition, we suppose there is a trusted functionality machine $\mathcal{F}_t$ that can correctly conduct all computations involved in \sysname, such as random value generation and HE encryption.

\vspace{0.1cm}
\noindent
\textbf{Secure RF Construction.}
In this stage, two protocols are involved, i.e., $\pi = \{$\protocol{FLE-Expan}$, $\protocol{FOS-Find}$\}$.
Since the two protocols are not independent from each other, we regard them as one protocol in the security analysis.
We use the following two theorems to show that there exist two independent simulators $\xi_{UD}$, $\xi_{CS}$ that can generate computationally indistinguishable views for $\pi$.
Since $CC$ does not participate in the computation of secure RF construction, it is trivial to analyse $\xi_{CC}$.

\begin{theorem}
\label{thm_train_adv_ud}
    For secure RF construction, there exists a PPT simulator $\xi_{UD}$ that can simulate an ideal view which is computationally indistinguishable from the real view of $\mathcal{A}_{UD}$.
\end{theorem}

\noindent \textit{proof.}
\quad
    Consider the scenario that a subset of \entity{UD} is corrupted.
    For the corrupted participants, $\xi_{UD}$ controls their local data and outputs. Therefore, $\xi_{UD}$ can simply follow the steps of $\pi$ with the local data.
    On behalf of the honest participants of \entity{UD}, $\xi_{UD}$ runs $\mathcal{A}_{UD}$ and uses dummy values to simulate its view.
    Specifically, the message that a participant $u\in $ \entity{UD} sends to \entity{CS} can only be a split vector set $\mathcal{W}_{\tau}$ or the best split encrypted with \func{HoEnc}.
    $\xi_{UD}$ can simulate these messages as follows.
    First, run $\mathcal{F}_t$ to generate random values to serve as dummy protocol inputs.
    Then, according to $\pi$, randomly enumerate several values from these dummy inputs for generating candidate splits.
    Finally, based on the candidate splits and dummy inputs, $\xi_{UD}$ obtains a simulate $\mathcal{W}_{\tau}$.
    Since the feature data of each participant never leaves the local database and $\mathcal{W}_{\tau}$ is derived from a random range, $\mathcal{A}_{UD}$ cannot identify whether $\mathcal{W}_{\tau}$ is dummy or not.
    Consequently, the simulated $\mathcal{W}_{\tau}$ is indistinguishable from a real split matrix.
    If chosen as the best split provider by \entity{CS}, $\xi_{UD}$ asks $\mathcal{F}_t$ to encrypt the chosen candidate split with \func{HoEnc}.
    According to Lemma~\ref{lem_enc_security}, the output is indistinguishable from an encrypted real value.
    In conclusion, such a $\xi_{UD}$ described in Theorem~\ref{thm_train_adv_ud} exists.
\QEDB

\begin{theorem}
\label{thm_train_adv_server}
    For secure RF construction, there exists a PPT simulator $\xi_{CS}$ that can simulate an ideal view which is computationally indistinguishable from the real view of $\mathcal{A}_{CS}$.
\end{theorem}

\noindent \textit{proof.}
\quad
    Similar to the proof of Theorem~\ref{thm_train_adv_ud}, for a corrupted \entity{CS}, $\xi_{CS}$ simply runs it by using its local data.
    For an ``honest'' \entity{CS}, $\xi_{CS}$ also uses the dummy values to simulate it for $\mathcal{A}_{CS}$.
    In this stage, the message sent by \entity{CS} is the sample selection vector $\vec{\mu}$, the feature selection vector $\vec{v}$ and the command to ask for best split.
    $\vec{v}$ is randomly selected to select a subset of features for current iteration of training.
    To operate the update of $\vec{\mu}$ and simulate the best split command, $\xi_{CS}$ first asks $\mathcal{F}_t$ to randomly generate dummy ground truths as protocol inputs.
    Then, based on the received split vectors from \entity{UD}, $\xi_{CS}$ asks $\mathcal{F}_t$ to compute the MSE or Gini coefficient with the dummy inputs and selects the best split vector.
    Since the returned command for uploading encrypted best split does not contain any special information, a simulated command or selection vector $\vec{v}$ is indistinguishable from a real one.
\QEDB

According to our security model, there are totally five attack types for \sysname, which are corrupted \entity{UD}, corrupted \entity{CS}, corrupted \entity{CC}, corrupted \entity{UD} and \entity{CS} or corrupted \entity{UD} and \entity{CC}.
From the above two theorems, it is natural to derive that under the five attacks, there always exists a corresponding simulator whose ideal view is indistinguishable from the real view of $\mathcal{A}$.
Therefore, we conclude that the RF construction stage of \sysname is secure in the \textit{honest-but-curious} model.

\vspace{0.1cm}
\noindent
\textbf{Secure RF Prediction.}
Since the security proof of RF Prediction is similar to RF construction, we give the detailed proof in Appendix~\ref{appendix_sec_prediction}.

\vspace{0.1cm}
\noindent
\textbf{Secure Participant Revocation.} 
For participant revocation, in addition to avoid the private data of the revoked participant to be revealed to the adversary, we also have to ensure that the revoked data are unavailable for the remaining participants.
Therefore, in the subsequent security analysis, we first prove that the \protocol{RFR-Revoc} is secure according to Definition~\ref{def_security}.
Then, we discuss how \protocol{RFR-Revoc} implements the second security goal.

In \protocol{RFR-Revoc}, only \entity{CS} and \entity{CC} participate in the computations.
The following two theorems state that there exist two independent simulators $\xi_{CS}$ and $\xi_{CC}$ that generate indistinguishable views for $\pi = \{$\protocol{FRF-Revoc}$\}$.

\begin{theorem}
\label{thm_revoc_cs}
    For secure participant revocation, there exists a PPT simulator $\xi_{CS}$ that can simulate an ideal view which is computationally indistinguishable from the real view of $\mathcal{A}_{CS}$.
\end{theorem}

\noindent \textit{proof.}
\quad
    For a corrupted \entity{CS}, $\xi_{CS}$ simply runs it with its local data.
    For a ``honest'' \entity{CS}, $\xi_{CS}$ first runs the simulator of \entity{CS} defined in Theorem~\ref{thm_train_adv_server} to simulate the protocol inputs of \protocol{FRF-Revoc}, i.e., a trained RF.
    Then, select the RDT nodes corresponding to the revoked participant from the simulated RF and send them to \entity{CC}.
    Finally, ask $\mathcal{F}_t$ to generate a random value and use it to refresh the ciphertexts returned from \entity{CC}.
    It can be found that the above exchanged messages only contain the encrypted splits of RDT nodes.
    Based on Lemma~\ref{lem_enc_security}, $\mathcal{A}_{CS}$ cannot distinguish whether the ciphertexts are simulated or not.
    Therefore, there exists a PPT simulator $\xi_{CS}$ that can simulate a view computationally indistinguishable from the real view of $\mathcal{A}_{CS}$.
\QEDB

\begin{theorem}
\label{thm_revoc_cc}
    For secure participant revocation, there exists a PPT simulator $\xi_{CC}$ that can simulate an ideal view which is computationally indistinguishable from the real view of $\mathcal{A}_{CC}$.
\end{theorem}

\noindent \textit{proof.}
\quad
    During participant revocation, \entity{CC} only takes on one task which is refreshing the ciphertexts.
    Therefore, no matter \entity{CC} is ``honest'' or not, $\xi_{CC}$ can simply simulate it by asking $\mathcal{F}_t$ to generate a random value and operate \func{HEncRef}.
    Also, it is easy to derive that the simulated view is indistinguishable from the real view of $\mathcal{A}_{CC}$.
\QEDB

From Theorem~\ref{thm_revoc_cs} and Theorem~\ref{thm_revoc_cc}, we prove that the participant revocation of \sysname does not leak any private data of the revoked participant.
Subsequently, we discuss that the revoked data are no longer available for the adversary.

In Section~\ref{subsub_participant_revocation}, we define two levels of revocation.
For the first-level revocation, we suppose that the revoked participant cannot be corrupted by $\mathcal{A}$.
For the second-level revocation, $\mathcal{A}$ is set to not have such restriction.
According to the design of \sysname, the data that a participant contributes to a learning federation in \sysname only contain the best split.
While utilizing the encrypted best split, \entity{CS} has to operate \func{HoLT} and the result is encrypted with both the \entity{CS} public key and its provider's public key.
If a participant is revoked and not corrupted, the only way to utilize its data is breaking the encryption algorithm \func{HoEnc}.
However, based on Lemma~\ref{lem_enc_security}, \func{HoEnc} is unresolvable in polynomial time.
Therefore, the first-level revocation can achieve its security goal, i.e., ``forward security''.
Additionally, to implement second-level revocation, we let \entity{CC} and \entity{CS} refresh the splits of revoked RDT nodes, respectively.
According to our security model, \entity{CC} and \entity{CS} cannot be simultaneously corrupted.
Thus, the refreshed splits are not available even $\mathcal{A}$ corrupts the revoked participant, which proves that \sysname achieves the ``backward'' security in the second-level revocation.

\section{Performance Evaluation}\label{sec_experiments}
In this section, we conduct experiments to prove the following three aspects:
The participant revocation does not obviously influence \sysname effectiveness (Section \ref{subsub_participant_revocation}).
\sysname is as effective as the other privacy-preserving RF frameworks (Section \ref{subsub_performance_comparison}).
\sysname is efficient for RF construction and prediction (Section \ref{sub_efficiency_eval}).

\textbf{Experiment Preparation.}
We use eight datasets from the UCI machine learning repository \cite{bache2013uci} in our experiments, four for classification and four for regression, shown in Table~\ref{table_dataset}.
For DP-PKC, we set $N = 1024$ to achieve 80-bits security level.
The default maximum RDT number and tree depth are set to $100$ and $10$, respectively.
Our experiments are performed with two laptops, one with an Intel Core i7-8565U CPU @1.8Ghz and 16G RAM and the other with an Intel Core i5-7200 CPU @2.50GHz and 8GB RAM.
The programs are written in Java.

\subsection{Effectiveness Evaluation}
To assess the effectiveness of \sysname, we first experiment with the performance of \sysname on eight datasets with different numbers of revoked participants.
Then, we compare the performance of \sysname with other privacy-preserving RF frameworks.

\subsubsection{Effectiveness Evaluation with Participant Revocation}
\label{subsub_participant_revocation}
The experiment results are presented in Table~\ref{table_classification_revocation} and Table~\ref{table_regression_revocation}.
For classification, we evaluate the impact of participant revocation on accuracy (ACC), recall rate (RR) and F1-score (F1).
For regression, we evaluate the impact of participant revocation on mean square error (MSE), mean absolute error (MAE) and R-Square (R2).
The six indicators are commonly used to assess the performance of a machine learning model \cite{zhang2017relationship, sam2017switching}.
Specially, we still suppose that each participant only owns one dimension of data, i.e. one feature.

\begin{figure}[htbp]
\centering
\subfigure[ACC with different numbers of RDT for Adult Income]{
\begin{minipage}[t]{0.47\linewidth}\label{fig_acc_adult}
\centering
\includegraphics[scale=0.18]{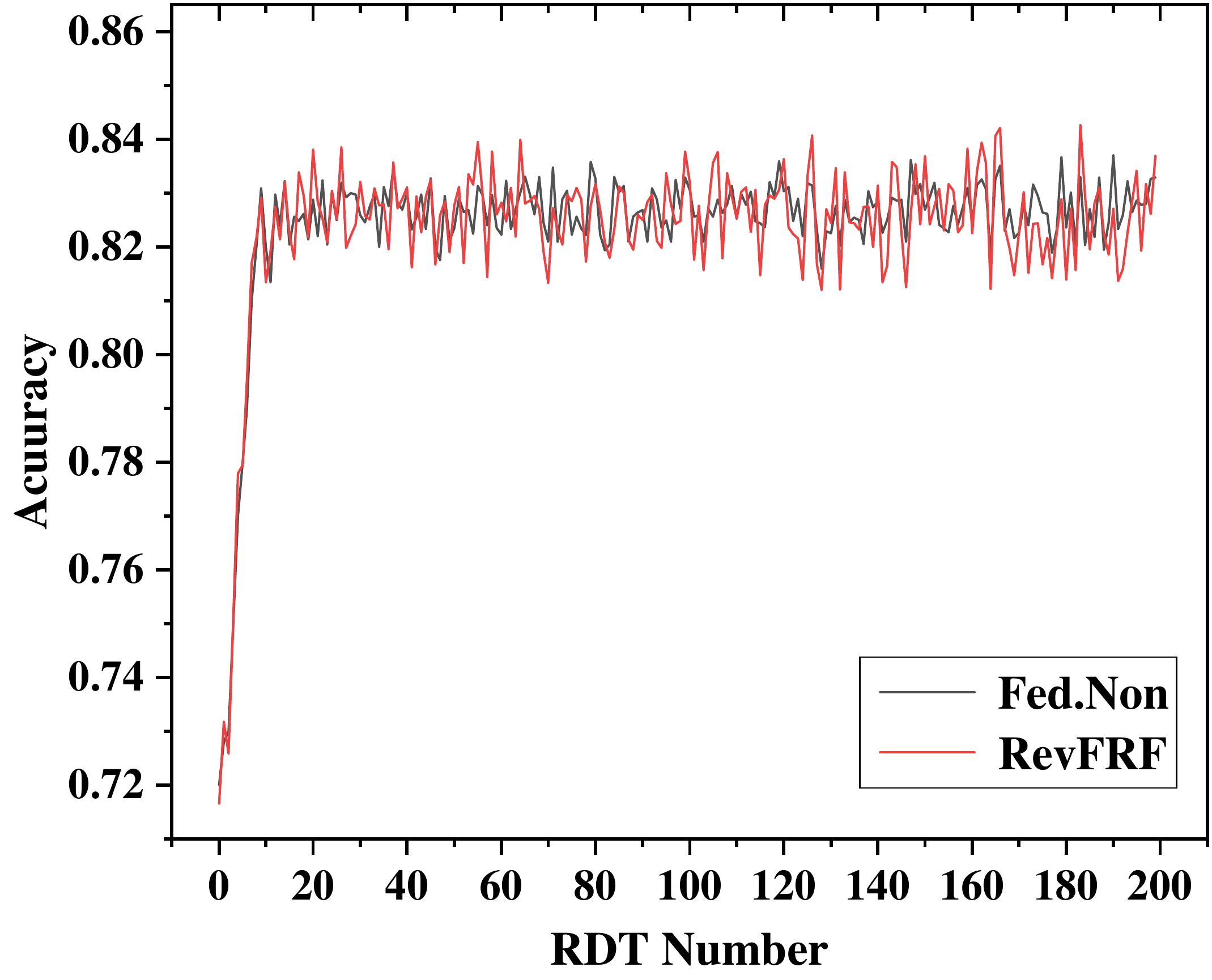}
\end{minipage}
}%
\hfill
\subfigure[ACC with different numbers of RDT for Bank Market]{
\begin{minipage}[t]{0.47\linewidth}\label{fig_acc_bank}
\centering
\includegraphics[scale=0.18]{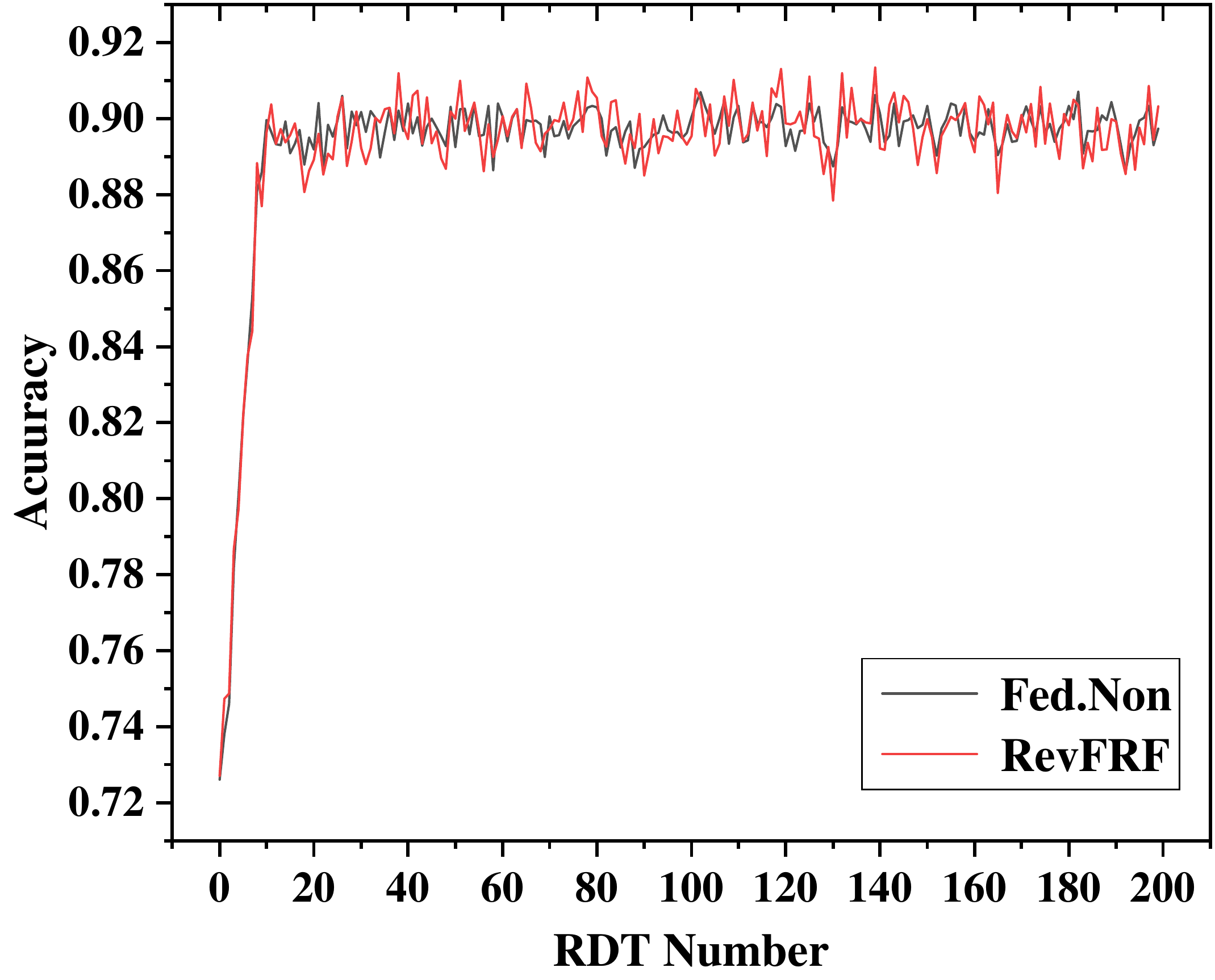}
\end{minipage}
}%
\caption{Classification Performance Evaluation (ACC)}
\label{fig_cla_acc}
\end{figure}

\begin{figure}[htbp]
\centering
\subfigure[Regression MSE with different numbers of RDT for Super Conduct]{
\begin{minipage}[t]{0.47\linewidth}\label{fig_acc_super}
\centering
\includegraphics[scale=0.18]{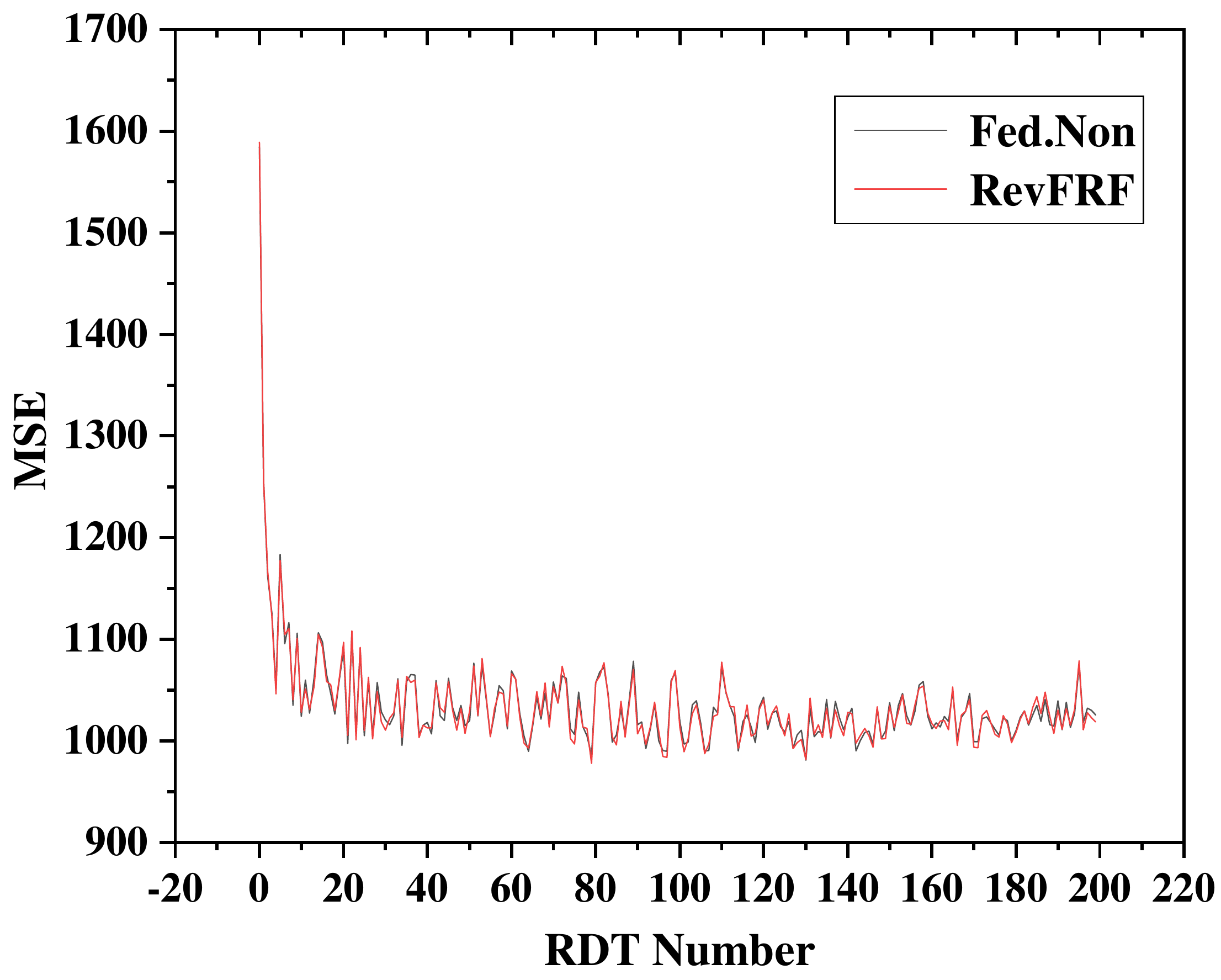}
\end{minipage}
}%
\hfill
\subfigure[Classification accuracy with different numbers of RDT for Appliance Energy]{
\begin{minipage}[t]{0.47\linewidth}\label{fig_acc_appliance}
\centering
\includegraphics[scale=0.18]{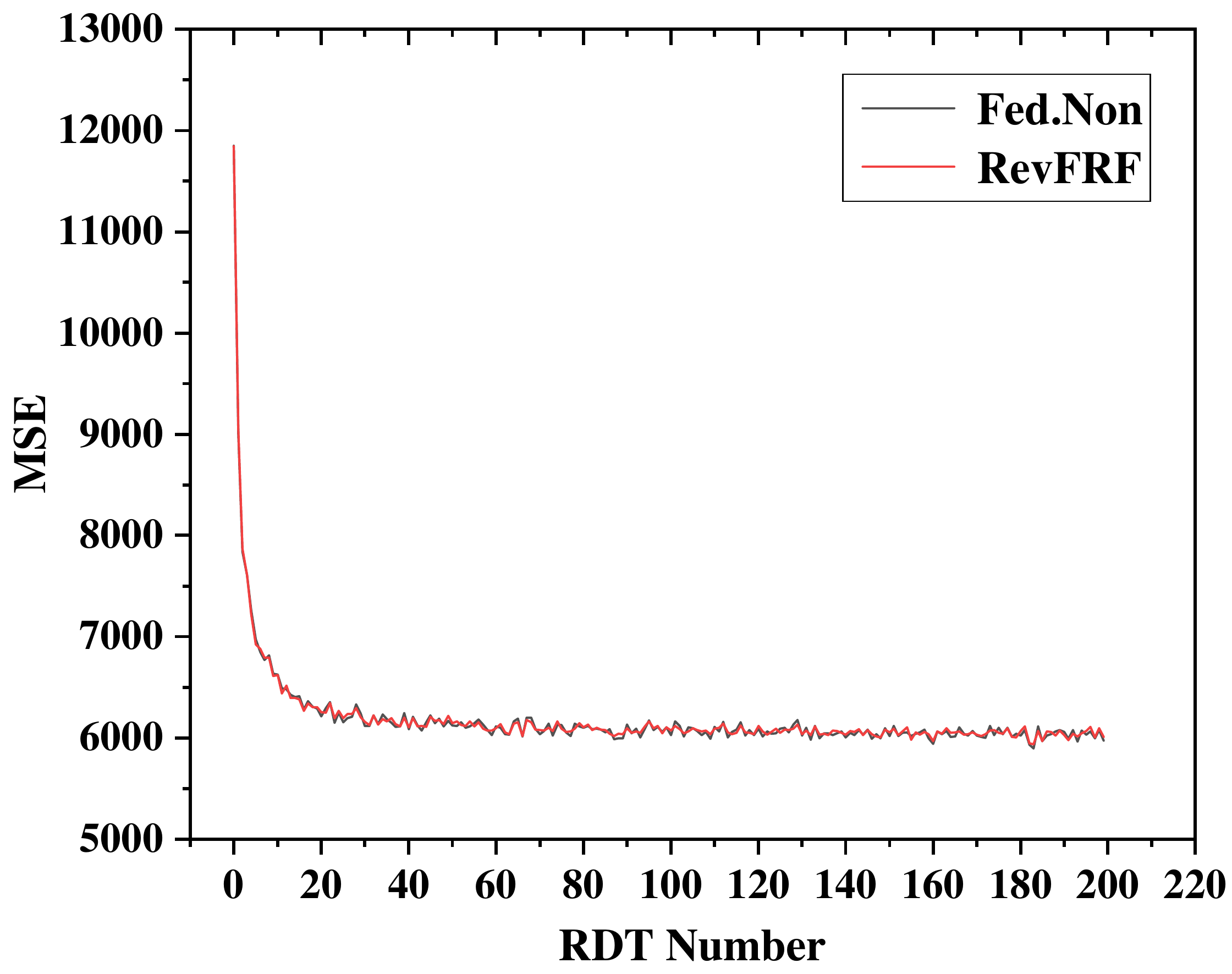}
\end{minipage}
}%
\caption{Regression Performance Evaluation (MSE)}
\label{fig_reg_mse}
\end{figure}

From the experiment results, the increase of revoked participants has little impact on the performance of \sysname when the percentage of revoked participants is less than 50\%.
Specifically, with maximum five revoked participants, the classification accuracy loss is less than 5\%, and the increased prediction error for regression is less than 3\%.
Therefore, in most cases, the revocation of parts of participants in \sysname hardly influences the effectiveness of \sysname.
Table~\ref{table_classification_revocation} and Table~\ref{table_regression_revocation} show that \sysname has a very small computation error compared to the original RF (Non.Fed).
The error is mainly caused by two aspects.
First, the computations of RF involve some random operations, which may lead to the uncontrollable differences in trained models.
Second, the data in parts of the datasets are stored in floating-point format.
As mentioned in Section~\ref{sec_premilinary}, we use the fixed-point format to represent these data to guarantee the security of HE.
Sometimes, the format transformation causes precision loss.
Fig.~\ref{fig_cla_acc} and Fig.~\ref{fig_reg_mse} further show the training process of the original RF framework (Non.Fed) and \sysname.
With the increased RDT number, \sysname attains approximate performance with the Non.Fed framework on two most important indicators for classification and regression, ACC and MSE.


\begin{table*}[!htbp]
\centering
\caption{Classification with Different Numbers of Revoked Participant}
\begin{tabular}{|c|c|c|c|c|c|c|c|c|c|}

\hline
\multicolumn{2}{|c|}{\multirow{2}{*}{Revoked Participants}} & \multicolumn{2}{c|}{Adult Income} & \multicolumn{2}{c|}{Bank Market} & \multicolumn{2}{c|}{Drug Consumption} & \multicolumn{2}{c|}{Wine Quality} \\

\cline{3-10}
\multicolumn{2}{|c|}{} & Non.Fed & \sysname & Non.Fed & \sysname & Non.Fed & \sysname & Non.Fed & \multicolumn{1}{c|}{\sysname} \\

\hline
\multirow{3}{*}{\xspace \xspace \xspace 0 \xspace \xspace \xspace} & ACC  & 85.63 & 85.63 & 90.97 & 90.77 & 90.62 & 90.53 & 68.52 & 68.51 \\

\cline{2-10}
& RR  & 86.17 & 86.15 & 91.69 & 91.46 & 91.27 & 91.18 & 66.27 & 66.17 \\

\cline{2-10}
& F1  & 85.69 & 85.72 & 91.19 & 91.01 & 90.55 & 90.59 & 64.64 & 64.37 \\


\hline
\multirow{3}{*}{1} & ACC  & 85.37 & 85.28 & 90.32 & 90.83 & 90.49 & 90.62 & 67.69 & 67.43 \\

\cline{2-10}
& RR  & 85.91 & 85.81 & 91.27 & 91.47 & 91.17 & 91.27 & 65.71 & 65.45 \\

\cline{2-10}
& F1  & 85.45 & 85.39 & 90.55 & 91.06 & 90.48 & 90.66 & 64.17 & 64.23 \\


\hline
\multirow{3}{*}{2} & ACC  & 85.11 & 84.43 & 90.38 & 90.49 & 90.28 & 90.44 & 67.41 & 67.38 \\

\cline{2-10}
& RR  & 85.61 & 85.02 & 91.29 & 91.23 & 90.97 & 91.09 & 65.66 & 64.68 \\

\cline{2-10}
& F1  & 85.2 & 84.56 & 90.61 & 90.73 & 90.29 & 90.52 & 64.2 & 63.51 \\


\hline
\multirow{3}{*}{3} & ACC  & 84.42 & 83.91 & 90.34 & 89.84 & 90.14 & 90.62 & 66.05 & 65.16 \\

\cline{2-10}
& RR  & 85.03 & 84.52 & 91.19 & 90.83 & 90.87 & 91.25 & 65.08 & 64.33 \\

\cline{2-10}
& F1  & 84.54 & 84.04 & 90.62 & 90.12 & 90.17 & 90.67 & 63.57 & 63.04 \\


\hline
\multirow{3}{*}{4} & ACC  & 83.53 & 83.41 & 90.26 & 90.16 & 90.11 & 90.33 & 65.52 & 65.33 \\

\cline{2-10}
& RR  & 84.19 & 84.06 & 91.17 & 91.01 & 90.87 & 90.89 & 63.83 & 63.66 \\

\cline{2-10}
& F1  & 83.67 & 83.54 & 90.51 & 90.43 & 90.09 & 90.44 & 62.35 & 62.44 \\


\hline
\multirow{3}{*}{5} & ACC  & 83.15 & 83.44 & 90.17 & 90.61 & 90.09 & 89.97 & 64.49 & 64.49 \\

\cline{2-10}
& RR  & 83.87 & 84.09 & 91.1 & 91.27 & 90.86 & 90.74 & 63.31 & 62.41 \\

\cline{2-10}
& F1  & 83.31 & 83.59 & 90.42 & 90.86 & 90.07 & 90.02 & 61.88 & 61.19 \\

\hline
\end{tabular}
\label{table_classification_revocation}
\end{table*}

\begin{table*}[!htbp]
\centering
\caption{Regression with Different Numbers of Revoked Participant}
\begin{tabular}{|c|c|c|c|c|c|c|c|c|c|}

\hline
\multicolumn{2}{|c|}{\multirow{2}{*}{Revoked Participant}} & \multicolumn{2}{c|}{Super Conduct} & \multicolumn{2}{c|}{Appliance Energy} & \multicolumn{2}{c|}{Insurance Company} & \multicolumn{2}{c|}{News Popularity} \\

\cline{3-10}
\multicolumn{2}{|c|}{} & Non.Fed & \sysname & Non.Fed & \sysname & Non.Fed & \sysname & Non.Fed & \multicolumn{1}{c|}{\sysname} \\

\hline
\multirow{3}{*}{\xspace \xspace \xspace 0 \xspace \xspace \xspace} & MSE & 994.69 & 1005.21 & 6038.64 & 6058.42 & 0.065 & 0.065 & 1414.85 & 1433.06 \\

\cline{2-10}
& MAE & 27.01 & 27.05 & 36.37 & 36.49 & 0.11 & 0.11 & 4.404 & 4.48 \\

\cline{2-10}
& R2 & 0.22 & 0.23 & 0.46 & 0.46 & 0.098 & 0.099 & 0.23 & 0.25\\


\hline
\multirow{3}{*}{1} & MSE & 995.24 & 1009.98 & 6052.24 & 6057.77 & 0.065 & 0.65 & 1424.79 & 1442.01 \\

\cline{2-10}
& MAE & 26.97 & 27.07 & 36.46 & 36.52 & 0.11 & 0.11 & 4.471 & 4.49 \\

\cline{2-10}
& R2 & 0.22 & 0.23 & 0.46 & 0.46 & 0.098 & 0.097 & 0.239 & 0.25 \\


\hline
\multirow{3}{*}{2} & MSE & 1000.7 & 1012.98 & 6063.66 & 6067.67 & 0.065 & 0.065 & 1423.34 & 1439.91 \\

\cline{2-10}
& MAE & 27.04 & 27.13 & 36.62 & 36.57 & 0.11 & 0.11 & 4.47 & 4.476 \\

\cline{2-10}
& R2 & 0.22 & 0.24 & 0.46 & 0.46 & 0.098 & 0.098 & 0.24 & 0.25 \\


\hline
\multirow{3}{*}{3} & MSE & 1010.79 & 1018.59 & 6078.93 & 6081.46 & 0.065 & 0.065 & 1430.98 & 1443.31 \\

\cline{2-10}
& MAE & 27.14 & 27.27 & 36.59 & 36.61 & 0.11 & 0.11 & 4.498 & 44.89 \\

\cline{2-10}
& R2 & 0.24 & 0.25 & 0.46 & 0.46 & 0.099 & 0.099 & 0.244 & 0.26 \\


\hline
\multirow{3}{*}{4} & MSE & 1012.88 & 1021.33 & 6090.12 & 6112.55 & 0.065 & 0.065 & 1441.42 & 1449.98 \\

\cline{2-10}
& MAE & 27.24 & 27.38 & 36.73 & 37.39 & 0.11 & 0.11 & 4.506 & 44.93 \\

\cline{2-10}
& R2 & 0.24 & 0.25 & 0.46 & 0.46 & 0.099 & 0.099 & 0.253 & 0.26 \\


\hline
\multirow{3}{*}{5} & MSE & 1021.2 & 1029.26 & 6135.74 & 6178.1 & 0.065 & 0.065 & 1445.43 & 1456.81 \\

\cline{2-10}
& MAE & 27.33 & 27.43 & 36.735 & 37.01 & 0.11 & 0.11 & 4.488 & 4.51 \\

\cline{2-10}
& R2 & 0.25 & 0.26 & 0.45 & 0.45 & 0.099 & 0.099 & 0.256 & 0.27 \\

\hline
\end{tabular}
\label{table_regression_revocation}
\end{table*}

\subsubsection{Effectiveness Comparison}
\label{subsub_performance_comparison}
Furthermore, we compare our performance against the protocols in \cite{rana2015differentially, vaidya2013random} on four datasets.
\cite{rana2015differentially} and \cite{vaidya2013random} are two of the most influential works on both secure RF construction and prediction.
Regard the original RF as the benchmark.
Our results illustrated in Table~\ref{table_performance_com} show that the introduction of noise in the DP based framework \cite{rana2015differentially} causes more obvious performance loss than \sysname with 100 RDTs.
Both \cite{vaidya2013random} and \sysname can achieve similar performance to the original RF.
Nevertheless, \sysname is designed to be more adaptive to handle the participant revocation condition, and require less overhead for RF construction (discussed in Section~\ref{sub_efficiency_eval}).
\begin{table}[!htbp]
\centering
\caption{Performance Comparison}
\begin{tabular}{|c|c|c|c|c|c|}

\cline{1-6}
\multicolumn{2}{|c|}{} & Non.Fed & \cite{rana2015differentially} & \cite{vaidya2013random} & \sysname \\

\hline
\multicolumn{1}{|c|}{\multirow{2}{*}{Cla.}} & Adult & 85.63 & 78.87 & 85.25 & 85.26 \\

\cline{2-6}
& Bank & 90.97 & 86.55 & 90.5 & 90.77 \\

\hline
\multicolumn{1}{|c|}{\multirow{2}{*}{Reg.}} & Conduct & 27.01 & 35.75 & 27.17 & 27.05 \\

\cline{2-6}
& Bike & 11.52 & 20.23 & 11.6 & 11.55 \\

\hline
\multicolumn{6}{c}{Cla. $\to$ Classification task evaluated with accuracy;}\\

\multicolumn{6}{c}{Reg. $\to$ Regression task evaluated with mean square error.} \\

\end{tabular}
\label{table_performance_com}
\end{table}

\subsection{Efficiency Evaluation}\label{sub_efficiency_eval}
To assess the efficiency of \sysname, we first theoretically analyse the computation and communication cost of \sysname.
Then, we compare the cost of \sysname with the existing frameworks.
Finally, we show the detailed cost analysis of secure RF construction and prediction stages, respectively.

\subsubsection{Theoretical Analysis}
The theoretical computation and communication cost of \sysname are given below.
Here, $\mathcal{O}(\cdot)$ is used to express the upper bound of computation and communication cost.

\begin{table*}[!htbp]
\centering
\caption{Comparison of Theoretical Computation Cost}
\begin{tabular}{|c|c|c|c|c|}

\hline
\multicolumn{3}{|c|}{} & \multicolumn{1}{c|}{Center Server} & \multicolumn{1}{c|}{Normal Participant}  \\

\hline
\multirow{3}{*}{Our} & \multirow{3}{*}{Comp.} & TC & $\mathcal{O}(t_{max} \cdot 2^{d_{max}} \cdot n_{f} \cdot \varrho)$ & $4.5 \mathcal{N} \cdot \mathcal{O}(t_{max}\cdot 2^{d_{max}})$ \\

\cline{3-5}
& & TP & $42\mathcal{N}\cdot \mathcal{O}(t_{max}\cdot d_{max})$ & $4.5 \mathcal{N}\cdot \mathcal{O}(t_{max}\cdot d_{max})$    \\

\cline{3-5}
& & PR & $3\mathcal{N}\cdot d_{max}\cdot n_d \cdot n_f \cdot \varrho$ & $4.5 \mathcal{N} \cdot d_{max} \cdot n_d$   \\







\hline
\multirow{2}{*}{\cite{vaidya2013random}} & \multirow{2}{*}{Comp.} & TC & N.A. & $1.5 \mathcal{N} \cdot \mathcal{O}(t_{max}\cdot 2^{d_{max}}\cdot n_t)$ \\

\cline{3-5}
& & TP & N.A. & $1.5 \mathcal{N}\cdot \mathcal{O}(t_{max}\cdot d_{max})$    \\



\hline
\multirow{1}{*}{\cite{bost2015machine}} & \multirow{1}{*}{Comp.} & TP & $7.5\mathcal{N}\cdot\mathcal{O}(t_{max}\cdot d_{max}\cdot\mathcal{N}) + 1.5\mathcal{N}\cdot\mathcal{O}(t_{max}\cdot d_{max})$ & $4.5\mathcal{N}\cdot\mathcal{O}(t_{max}\cdot d_{max}\cdot\mathcal{N}) + 3\mathcal{N}\cdot\mathcal{O}(t_{max}\cdot d_{max})$ \\


\hline
\multirow{1}{*}{\cite{wu2016privately}} & \multirow{1}{*}{Comp.} & TP & $6\mathcal{N}\cdot\mathcal{O}(t_{max}\cdot d_{max}\cdot\mathcal{N})$ & $3\mathcal{N}\cdot\mathcal{O}(t_{max}\cdot d_{max}\cdot\mathcal{N})$ \\


\hline
\multicolumn{5}{c}{Comp. $\to$ Computation Cost; Comm. $\to$ Communication Cost; TC $\to$ Tree Construction; TP $\to$ Tree Prediction $n_t\gg n_f$.}\\

\end{tabular}
\label{table_theoretical}
\end{table*}

\vspace{0.1cm}
\noindent
\textbf{Computation Cost.}
Assume a regular exponentiation operation requires $1.5\mathcal{N}$ multiplications \cite{knuth2014art}.
$\mathcal{N}$ is the data length of the exponent.
Since the cost of an exponentiation operation is much more than an addition or multiplication operation, the fixed numbers of addition and multiplication operations are ignored in our analysis.
Thus, we can conclude that \func{HoEnc} requires $3\mathcal{N}$ to encrypt a message.
\func{HEncRef} needs $1.5\mathcal{N}$ multiplications.
\func{ParHDec1} and \func{ParHDec2} totally need $3\mathcal{N}$ multiplications.
\func{HoLT} requires $39\mathcal{N}$ multiplications.

Represent the maximum tree number, the maximum tree depth, the randomly selected feature number and the candidate splits provided by a participant as $t_{max}$, $d_{max}$, $n_{f}$ and $\varrho$ respectively.
In secure RF construction, \sysname generates $\mathcal{O}(t_{max}\cdot 2^{d_{max}})$ RDT nodes.
For each node, \entity{CS} computes $n_{f}$ MSE or Gini functions.
The selected normal participant conducts one \func{HoEnc}.
Thus, the computation cost of \entity{CS} in this stage is $ \mathcal{O}(t_{max} \cdot 2^{d_{max}} \cdot n_{f} \cdot \varrho)$ multiplications.
The normal participant takes $4.5 \mathcal{N} \cdot \mathcal{O}(t_{max}\cdot 2^{d_{max}})$ multiplications.

In secure RF prediction, \sysname computes $\mathcal{O}(t_{max}\cdot d_{max})$ RDT nodes.
For each node, \entity{CS} conducts one \func{HoLT} and one \func{ParHDec2}. 
The selected normal participant conducts one \func{ParHDec1}.
Therefore, \entity{CS} requires $42 \mathcal{N}\cdot \mathcal{O}(t_{max}\cdot d_{max})$ multiplications.
And the normal participant computes up to $4.5 \mathcal{N}\cdot \mathcal{O}(t_{max}\cdot d_{max})$ multiplications.

For the worst case of participant revocation, the computation cost is the same as RF construction.
However, the expected cost of RF rebuilding in participant revocation is much less than RF construction.
Suppose the probability for an RDT node to be revoked in an RDT is identical and the number of revoked participants is $n_d$.
The expected number of revoked RDT nodes for an RDT is $\frac{n_d}{2^{d_{max}}}\sum_{i = 0}^{d_{max} - 1} 2^{d_{max} - i} \cdot 2^i = d_{max} \cdot n_d$.
Thus, for the first-level revocation, the expected computation costs of \entity{CS} and normal participant are $d_{max} \cdot n_d \cdot n_{f} \cdot \varrho$ and $4.5 \mathcal{N} \cdot d_{max} \cdot n_d$, respectively.
For the second-level revocation, the expected computation cost \entity{CS} is $3\mathcal{N}\cdot d_{max} \cdot n_d \cdot n_{f} \cdot \varrho$, and the cost of normal participant is the same as the first-level revocation.

\vspace{0.1cm}
\noindent
\textbf{Communication Cost.}
Assume that the output of \func{HoEnc} is $4\mathcal{N}$ bits and the number of total samples are $n_t$.
Similarly, we derive the communication cost as follows.

In secure RF construction, \sysname costs $(4\mathcal{N} + n_f\cdot n_t \cdot \varrho) \cdot \mathcal{O}(t_{max}\cdot 2^{d_{max}})$ bits.
In secure RF prediction, \sysname costs $28\mathcal{N}\cdot \mathcal{O}(t_{max}\cdot d_{max})$ bits.
Similar to computation cost, we give the expected communication cost of participant revocation.
For the first-level revocation, \sysname costs $(4\mathcal{N} + n_f\cdot n_t \cdot \varrho) \cdot d_{max}\cdot n_d$ bits.
For the second-level revocation, \sysname costs $(8\mathcal{N} + n_f\cdot n_t \cdot \varrho) \cdot d_max\cdot n_d$ bits.

\subsubsection{Cost Comparison}
We compare the \sysname efficiency against the protocols in \cite{bost2015machine, wu2016privately, vaidya2013random}.
\cite{rana2015differentially} is ignored because of its performance loss.
Specially, \cite{bost2015machine} and \cite{wu2016privately} are designed for RF prediction and cannot be used for RF construction.
\cite{vaidya2013random} does not have a \entity{CS}.
For tree construction, we calculate the computation and communication cost of generating 100 RDT nodes with 5000 samples of the Adult Income dataset.
For tree prediction, the computation and communication cost is evaluated with ten trees whose depth is ten.
The results reported in the experiments are an average over 10 trials.
Table~\ref{table_overhead} illustrates the comparison results.
Compared to \cite{bost2015machine} and \cite{wu2016privately}, \sysname outperforms them in terms of computation and communication cost.
Although \cite{rana2015differentially} costs less in the prediction stage, \sysname is over 2000 times faster than it in the tree construction stage.
The reason is that \cite{rana2015differentially} needs to encrypt the whole dataset for secure RF construct with a HE algorithm, which is not very practical in real-world applications.
On the contrary, we minimize the HE usage (once per RDT node) by virtue of federated learning.
The experiment results also obey the comparison result of theoretical computation cost, shown in Table~\ref{table_theoretical}.

Furthermore, Table~\ref{table_revocation_rebuilding} illustrates the time required for RF rebuilding.
In the experiment, we conduct \sysname with different numbers of revoked participants and \sysname without the participant revocation mechanism (i.e., Irev.Fed), which needs to rebuild the whole RF.
From the experiment result, we can derive that the computation cost for an irrevocable RF framework is greatly higher than \sysname.
And with more participants revoked from the learning federation, the time required for \sysname to rebuild a federation also increases.










\begin{table}[!htbp]
\centering
\caption{Efficiency Comparison}
\begin{tabular}{|c|c|c|c|c|c|}

\hline
\multicolumn{2}{|c|}{\multirow{2}{*}{}} & \multicolumn{2}{c|}{Computation (s)} & \multicolumn{2}{c|}{Communication (MB)} \\

\cline{3-6}
\multicolumn{2}{|c|}{} & TC & TP & TC & TP \\

\hline
\multirow{2}{*}{\cite{bost2015machine}} & \entity{CS} & N.A. & 28.56 & \multirow{2}{*}{N.A.} & \multirow{2}{*}{81.58} \\

\cline{2-4}
& \entity{UD} & N.A. & 57.15 &  &  \\
 
\hline
\multirow{2}{*}{\cite{wu2016privately}} & \entity{CS} & N.A. & 26.32 & \multirow{2}{*}{N.A.} & \multirow{2}{*}{9.71} \\

\cline{2-4}
& \entity{UD} & N.A. & 57.15 &  &  \\
 
\hline
\multirow{2}{*}{\cite{vaidya2013random}} & \multirow{2}{*}{\entity{UD}} & \multirow{2}{*}{3932.41}  & \multirow{2}{*}{0.55}  & \multirow{2}{*}{0.42}  & \multirow{2}{*}{0.41}  \\

&  &  &  &  & \\ 

\hline
\multirow{2}{*}{\sysname} & \entity{CS} & 3.09 & 13.33 & \multirow{2}{*}{1.07} & \multirow{2}{*}{0.78} \\

\cline{2-4}
& \entity{UD} & 2.27. & 0.08 & &  \\

\hline

\end{tabular}
\label{table_overhead}
\end{table}

\begin{table}[!htbp]
\centering
\caption{Rebuilding Time with Different Numbers of Revoked Participants}
\begin{tabular}{|p{2cm}|c|c|c|c|c|c|}

\hline
Revoked Participant & 1 & 2 & 3 & 4 & 5 & IRev.Fed \\

\hline
Revoked RDT Nodes & 16.1 & 37 & 50.1 & 83.1 & 136.3 & 432.5 \\

\hline
Rebuilding Time & 1.2 & 2.7 & 3.6 & 6.1 & 9.8 & 31.1 \\

\hline
\multicolumn{7}{c}{IRev.Fed $\gets$ The irrevocable federated RF framework} \\

\end{tabular}
\label{table_revocation_rebuilding}
\end{table}

\subsubsection{Efficiency Evaluation of \sysname}
\begin{figure}[htbp]
\centering
\subfigure[Computation and communication cost of secure RF construction]{
\begin{minipage}[t]{0.45\linewidth}\label{fig_tb_cost_evaluation}
\centering
\includegraphics[scale=0.17]{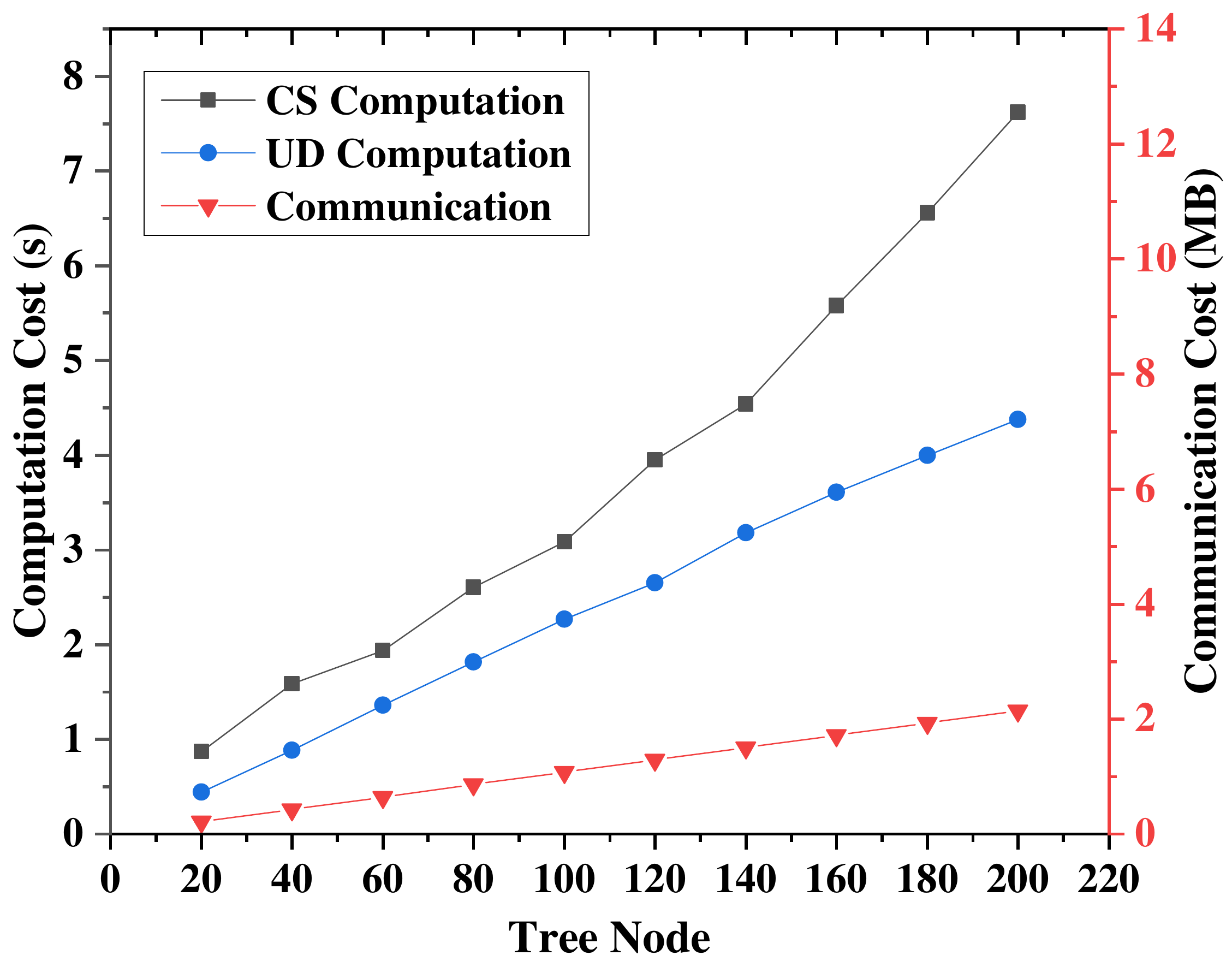}
\end{minipage}
}%
\hfill
\subfigure[Computation and communication cost of secure RF Prediction]{
\begin{minipage}[t]{0.45\linewidth}\label{fig_tp_cost_evaluation}
\centering
\includegraphics[scale=0.17]{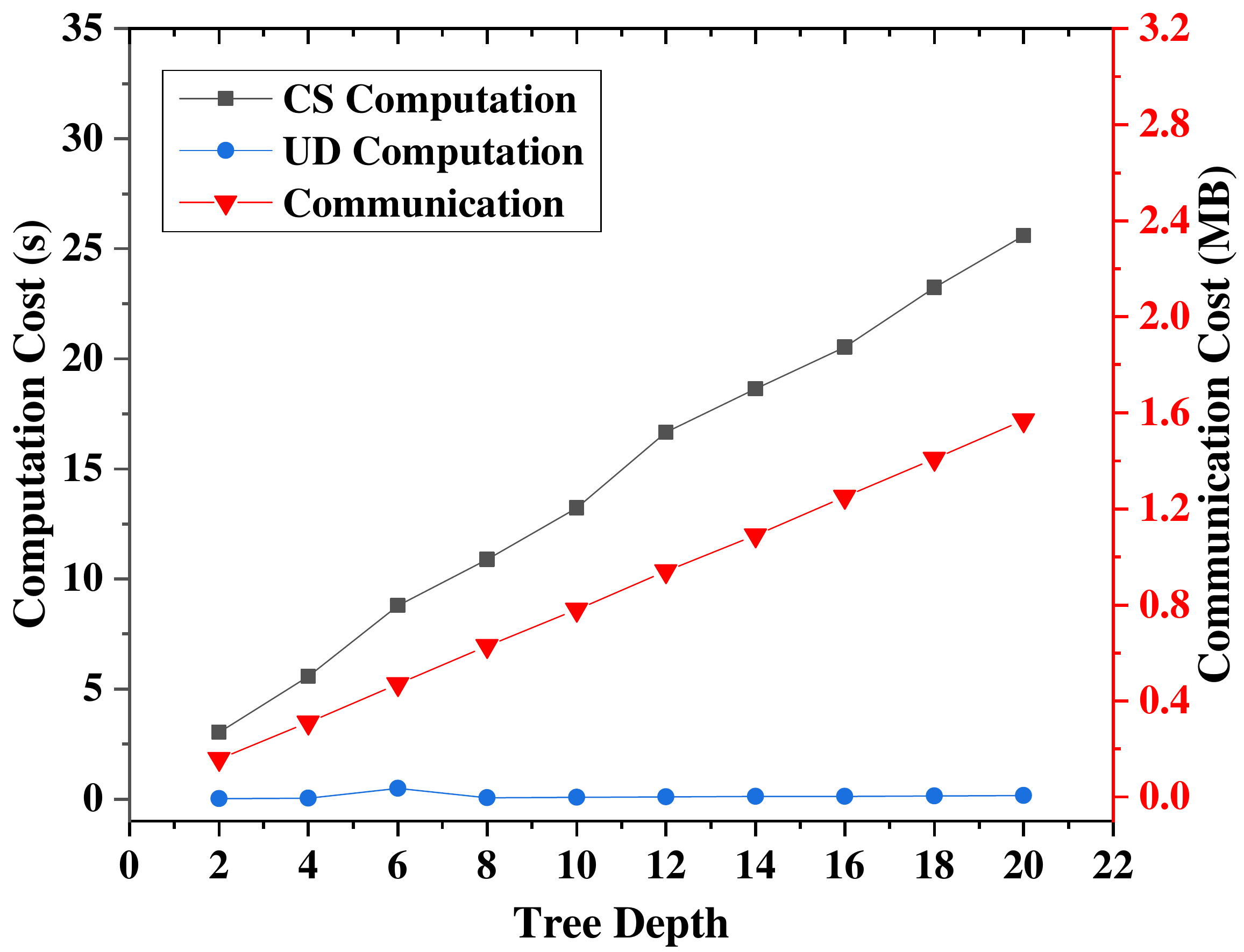}
\end{minipage}
}%
\caption{Computation and communication cost of \sysname}
\label{fig_reg_mse}
\end{figure}

Fig.~\ref{fig_tb_cost_evaluation} and Fig.~\ref{fig_tp_cost_evaluation} further show the overhead of secure RF construction and prediction under different numbers of RDT nodes and tree depth.
When the generated RDT nodes increase in the tree construction stage, both the computation cost and the communication cost of \sysname linearly grow.
Along with the increase of tree depth, the cost growth trend of tree prediction is analogous to tree construction, i.e. linear growth.
The experimental results accord with the theoretical analysis.


\section{Conclusion}\label{sec_conclusion}
In this paper, we introduced the revocable federated learning concept and proposed a revocable federated RF framework, \sysname.
In \sysname, we presented a suite of HE based secure protocols to achieve privacy-preserving RF construction and prediction.
Based on the specially designed RDT node storage method, \sysname also supported secure participant revocation.
Moreover, we gave a comprehensive analysis to show that \sysname could resolve all the security problems for a revocable federated learning framework.
The results of extensive experiments proved that compared to the existing privacy-preserving RF frameworks, our federated framework had lower performance loss and higher efficiency.


\bibliographystyle{ACM-Reference-Format}
\bibliography{references}

\appendix
\section{Homomorphic Encryption}
Here, we give more detailed definition abouts the fixed-point data format and the HE functions.

\subsection{Fixed-Point Data Format}\label{appendix_fixed_point}
In \sysname, we use the fixed-point format \cite{liuprivacyobj} to represent the input and output of HE functions.
In this format, we represent an arbitrary number $x\in \mathbb{R}$ to be $\hat{x} = \lfloor x \cdot 10 ^{c} \rfloor \mod N$, where $\lfloor\cdot \rfloor$ is a function to round a number to its nearest integer, $c$ is a fixed integer used to control representation precision.
For example, given $x = 3.1415$, $c = 2$ and $N = 13$, $x$ is represented as $\hat{x} = \lfloor 3.1415 \cdot 10 ^{2} \rfloor \mod 13 = 314 \mod 13$.
This data representation method can cause a little precision loss of precision but is essential to ensure the security of the HE functions.

\subsection{Homomorphic Encryption Functions}\label{appendix_functions}
The detailed definitions of the HE functions are given as follows.

\textbf{Encryption (\func{HoEnc}).}
To encrypt a plaintext message $m\in \mathbb{Z}_N$, the encipherer $u$ first selects a random number $r\in [1, N/4]$.
Then, compute $C_1 = pk^{r}_{u}(1 + mN) = g^{sk_{u}\cdot r} (1 + mN) \mod N^2$ and $C_2 = g^r \mod N^2$.
The final output of \func{HoEnc} is $[\![m]\!]_{pk_{u}} = \{C_1, C_2\}$.

\textbf{Re-Encryption (\func{HReEnc}).}
To re-encrypt a ciphertext $[\![m]\!]_{pk_{u}}$, \entity{CS} first computes $h_{2} = C_2^{sk_{CS}} = g^{r \cdot sk_{CS}} \mod N^2$.
Then, $CS$ updates $C_1' = C_1 \cdot h_2 = g^{(sk_{u} + sk_{CS})r}(1 + mN) \mod N^2$.
The final output of \func{HReEnc} is $[\![m]\!]_{pk_{\Sigma}} = \{C_1', C_2'\}$.

\textbf{Ciphertext Refresh (\func{HEncRef}).}
To refresh a ciphertext $[\![m]\!]_{pk_{u}}$, \entity{CS} or \entity{CC} first generates a random value $r_p$, and then, compute:
\begin{equation*}
    C_1' = C_1 \cdot C_2^{r_p} = g^{(sk_{u} + r_p)r}(1 + mN) \mod N^2.
\end{equation*}
The final output of \func{HEncRef} is $[\![m]\!]_{new} = \{C_1', C_2\}$.

\textbf{Partial Decryption (\func{ParHDec1} and \func{ParHDec2}).}
To decrypt a cihpertext $[\![m]\!]_{pk_{\Sigma}}$, $pk_{\Sigma} = pk_{CS} + pk_{u}$, we have to operate both \func{ParHDec1} and \func{ParHDec2}.
Suppose the decrypter is \entity{CS}.
For \func{ParHDec1}, \entity{CS} asks $u$ to update $C_1' = C_1 / C_2^{sk_{u}} = g^{sk_{CS}\cdot r}(1 + mN) \mod N^2$.
The output of \func{ParHDec1}, $[\![m]\!]_{pk_{CS}} = \{C_1', C_2\}$, is returned to \entity{CS}.
For \func{ParHDec2}, \entity{CS} computes $m = L(C_1' / C_2^{sk_{CS}} \mod N^2) \mod N$,
where $L(x) = (x - 1)/N$.
The final output of \func{ParHDec2} is $m$.

To implement \func{HoLT}, we still have to introduce a HE addition function (HoAdd).

\textbf{Addition across Different Domains(\func{HoAdd}).}
Given two ciphertext $[\![m_1]\!]_{pk_{u_1}}$ and $[\![m_2]\!]_{pk_{u_2}}$, \func{HoAdd} outputs the encrypted addition result $[\![m_1 + m_2]\!]_{pk_{\Sigma}}$.
\func{HoAdd} can only be conducted by \entity{CS}.
To achieve this, \entity{CS} first selects two random numbers, $\alpha_1, \alpha_2\in \mathbb{Z}_N$, and computes:
\begin{equation*}
\begin{aligned}
    &[\![m_1 + \alpha_1]\!]_{pk_{u_1}} = [\![m_1]\!]_{pk_{u_1}}\cdot [\![\alpha_1]\!]_{pk_{u_1}} = \{C_1', C_2'\}, \\
    &[\![m_2 + \alpha_2]\!]_{pk_{u_2}} = [\![m_2]\!]_{pk_{u_2}}\cdot [\![\alpha_2]\!]_{pk_{u_2}} = \{C_1'', C_2'\}.
\end{aligned}
\end{equation*}
Then, \entity{CS} uses its strong private key $\lambda_1$ to partially decrypt:
\begin{equation*}
\begin{aligned}
    &P1' = C_1'^{\lambda_1} = g^{r_1\cdot sk_{u_1}\cdot \lambda_1}(1 + (m_1 + \alpha_1)N\lambda_1)\mod N^2, \\
    &P1''= C_1''^{\lambda_1} = g^{r_2\cdot sk_{u_2}\cdot \lambda_1}(1 + (m_2 + \alpha_2)N\lambda_1)\mod N^2.
\end{aligned}
\end{equation*}
$C_1'$, $C_1''$, $P1'$ and $P1''$ are sent to \entity{CC}.
Next, \entity{CC} also uses its own strong private key $\lambda_2$ to partially decrypt the cipertexts:
\begin{equation*}
\begin{aligned}
    &P2'= C_1'^{\lambda_2} = g^{r_1\cdot sk_{u_1}\cdot \lambda_2}(1 + (m_1 + \alpha_1)N\lambda_2)\mod N^2, \\
    &P2''= C_1''^{\lambda_2} = g^{r_2\cdot sk_{u_2}\cdot \lambda_2}(1 + (m_2 + \alpha_2)N\lambda_2)\mod N^2.
\end{aligned}
\end{equation*}
\entity{CC} can subsequently obtain the addition result that are masked with two random values by calculating:
\begin{equation*}
\begin{aligned}
    m_1 + m_2 + \alpha_1 + \alpha_2 = L(P1' \cdot P2') + L(P1'' \cdot P2'') \mod N,
\end{aligned}
\end{equation*}
\begin{equation*}
\begin{aligned}
    P1' \cdot P2' &= g^{r_1\cdot sk_{u_1}\cdot \lambda}(1 + (m_1 + \alpha_1)N\lambda) \\
                  &= 1 + (m_1 + \alpha_1)N\lambda \mod N^2,
\end{aligned}
\end{equation*}
\begin{equation*}
\begin{aligned}
    P1'' \cdot P2'' &= g^{r_2\cdot sk_{u_2}\cdot \lambda}(1 + (m_2 + \alpha_2)N\lambda) \\
                    &= 1 + (m_2 + \alpha_2)N\lambda \mod N^2.
\end{aligned}
\end{equation*}
Express the masked addition result as $M = m_1 + m_2 + \alpha_1 + \alpha_2$.
\entity{CC} encrypts $[\![M]\!]_{pk_{\Sigma}} = \func{HoEnc}(pk_{u_1} + pk_{u_2}, M)$ with the public keys of $u_1$ and $u_2$ and returns $[\![M]\!]_{pk_{\Sigma}}$ to \entity{CS}.
As received the encrypted result, \entity{CS} can compute the final output $[\![m_1 + m_2]\!]_{pk_{\Sigma}} = [\![M]\!]_{pk_{\Sigma}}\cdot ([\![\alpha_1 + \alpha_2]\!]_{pk_{\Sigma}})^{N - 1}$.

\textbf{Secure Comparison (\func{HoLT}).}
Given two ciphertext $[\![m_1]\!]_{pk_{u_1}}$ and $[\![m_2]\!]_{pk_{u_2}}$, \func{HoLT} is used to judge their relationship, i.e., $m_1 < m_2$ or $m_1 \neq m_2$.
To achieve this goal, \entity{CS} first computes $[\![2m_1 + 1]\!]_{pk_{u_1}} = [\![m_1]\!]_{pk_{u_1}}^2 \cdot [\![m_1]\!]_{pk_{u_1}} \cdot [\![1]\!]_{pk_{u_1}}$ and $[\![2m_2]\!]_{pk_{u_2}} = [\![m_2]\!]_{pk_{u_2}}^2 \cdot [\![m_2]\!]_{pk_{u_2}}$.
Then, \entity{CS} randomly selects a number $\alpha$ from $\{0, 1\}$.
If $\alpha = 1$, \entity{CS} calculates $[\![\beta]\!]_{pk_{\Sigma}} = $\func{HoAdd}$([\![2m_1 + 1]\!]_{pk_{u_1}}, ([\![2m_2]\!]_{pk_{u_2}})^{N-1})$; otherwise, $[\![\beta]\!]_{pk_{\Sigma}} = $\func{HoAdd}$([\![2m_2]\!]_{pk_{u_2}}, ([\![2m_1 + 1]\!]_{pk_{u_1}})^{N-1})$.
\entity{CS} selects a random number $r$ satisfying $||r|| < ||N||/4$ and computes $[\![\beta']\!]_{pk_{\Sigma}} = ([\![\beta]\!]_{pk_{\Sigma}})^{r}$.
Next, \entity{CS} and \entity{CC} decrypt the $[\![\beta']\!]_{pk_{\Sigma}}$ with their strong private keys $\lambda_1$ and $\lambda_2$ in the same way as \func{HoAdd}.
If the decryption result $||\beta'|| > ||N|| / 2$, \entity{CC} denotes $l = 1$; otherwise, $l = 0$.
Subsequently, \entity{CC} sends $[\![l]\!]_{pk_{\Sigma'}} = $\func{HoEnc}$(pk_{CS} + pk_{u_2}, l)$ to \entity{CS}, where $pk_{\Sigma'} = pk^{CS} \cdot pk^{u_2}$.
Specially, \entity{CS} needs to update $[\![l]\!]_{pk_{\Sigma'}} = [\![1]\!]_{pk_{\Sigma'}}\cdot ([\![l]\!]_{pk_{\Sigma'}})^{N - 1}$ if $\alpha = 0$.
Finally, \entity{CS} obtains the comparison result $l$ by operating \func{ParHDec1} and \func{ParHDec2}.
If $l = 1$, $m_1$ is less than $m_2$; otherwise, $m_1 \geq m_2$.


\section{Quality Assessment Functions}\label{appendix_quality_assessment}
For regression, \entity{CS} computes the addition of mean squared errors for two child nodes.
\begin{equation}\label{eq_mse}
\begin{aligned}
    E(D_1, D_2, \vec{w}) = \frac{1}{n_1}\sum_{y_i\in D_1}(y_i - \bar{y}_{D_1})^2 +
    \frac{1}{n_2}\sum_{y_i\in D_2}(y_i - \bar{y}_{D_2})^2,
\end{aligned}
\end{equation}
where $D_1$ and $D_2$ are the samples of two child nodes obtained by the split vector $\vec{w}$;
$n_1$ and $n_2$ are the numbers of samples in $D_1$ and $D_2$;
$y_i$ is the ground-truth with index $i$;
$\bar{y}_{D_1}$ and $\bar{y}_{D_2}$ are the average values of the ground-truth in $D_1$ and $D_2$, respectively.
For classification, \entity{CS} computes the Gini coefficients.
\begin{equation}\label{eq_gini}
\begin{aligned}
    G(D_1, D_2, \vec{w}) = (1 - \sum_{x_i \in D_1}^{n_1}\sum_{j = 1}^{k} p_{x_i, j}) + (1 - \sum_{x_i \in D_2}^{n_2}\sum_{j = 1}^{k} p_{x_i, j}).
\end{aligned}
\end{equation}
$k$ is the number of classes.
$p_{x_i, j}$ is the probability of sample $x_i$ to be classified into the class $j$.

\section{Federated RF Prediction Protocols}\label{appendix_prediction}
The two types of RF prediction mentioned in Section~\ref{sub_rf_pre} are presented in Protocol~\ref{pro_frf_predict}, \ref{pro_ft_predict}, \ref{pro_frf_testing} and \ref{pro_ft_testing}, respectively.
\begin{algorithm}[ht]
  \caption{Federated  RF Prediction (\protocol{FRF-Predict})}
  \label{pro_frf_predict}
  \begin{algorithmic}[1]
    \Require
    A random forest $\mathcal{T}$;
    a prediction request $\vec{x} = \{x_1, x_2, ..., x_{\gamma}\}$, $\gamma = |\mathcal{F}|$;
    \Ensure
    The prediction result $O$.

    \State The requestor $u_0$ encrypts $[\![\vec{x}]\!]_{pk_0} = ([\![x_1]\!]_{pk_0}, [\![x_2]\!]_{pk_0}, ...,$ $ [\![x_{\gamma}]\!]_{pk_0}) = $\func{HoEnc}$(pk_0, \vec{x})$ and uploads it to \entity{CS}.
    


    \For{$T_i\in \mathcal{T}$}
        \State \entity{CS} computes $r_i = $\protocol{FT-Predict}$(T_i, 1, [\![\vec{x}]\!]_{pk_0})$ and $\mathcal{R} = \mathcal{R}\cup r_i$
    \EndFor

    \If{$task$ is $0$} \# regression task
        \State \entity{CS} computes the final output $O = \frac{1}{|\mathcal{R}|}\sum_{i = 1}^{|\mathcal{R}|} r_i$
    \Else \xspace \# classification task
        \State \entity{CS} counts $r_i \in \mathcal{R}$, and selects the one with most votes as the final classification result $O$.
    \EndIf

    \State \entity{CS} returns $O$ to $u_0$.
    


  \end{algorithmic}
\end{algorithm}

\begin{algorithm}[ht]
  \caption{Federated Tree Prediction (\protocol{FT-Predict})}
  \label{pro_ft_predict}
  \begin{algorithmic}[1]
    \Require
    A tree node $Node$;
    current tree depth $d_c$;
    an encrypted prediction request $[\![\vec{x}]\!]_{pk_0} = ([\![x_1]\!]_{pk_0}, [\![x_2]\!]_{pk_0}, ..., [\![x_{\gamma}]\!]_{pk_0})$.
    \Ensure
    The prediction result of a tree $r$.

    \If{$d_c \geq d_{max}$}
        \State \Return the weight of $Node$.
    \EndIf


    \State \entity{CS} extracts the split $[\![s_{\tau}]\!]_{pk_{\tau}}$ and the corresponding split provider $u_{\tau}$ from $Node$ and computes $[\![l]\!]_{pk_{\Sigma}} = $\func{HoLT}$([\![x_{\tau}]\!]_{pk_0}, [\![s_{\tau}]\!]_{pk_{\tau}})$, where $pk_{\Sigma} = pk_{CS} + pk_{\tau}$.

    \State \entity{CS} forwards the requestor ID and the half-decrypted ciphertext, $u_0, [\![l]\!]_{pk_{\Sigma}}$.

    \State $u_{\tau}$ returns $[\![l]\!]_{pk_{CS}} = $\func{ParHDec1}$(sk_{\tau}, [\![l]\!]_{pk_{\Sigma}})$ if he accepts the requestor.

    \State \entity{CS} decrypts $l = $\func{ParHDec2}$(sk_{CS}, [\![l]\!]_{pk_{CS}})$ and updates $d_c = d_c + 1$.
    If $l$ is $0$, operate \protocol{FT-Predict}$(Node\to rightChild,d_c, [\![\vec{x}]\!]_{pk_0})$; otherwise, \protocol{FT-Predict}$(Node\to leftChild, d_c, [\![\vec{x}]\!]_{pk_0})$.
    

  \end{algorithmic}
\end{algorithm}

\begin{algorithm}[ht]
  \caption{Federated RF Prediction for Testing Data (\protocol{FRF-Test})}
  \label{pro_frf_testing}
  \begin{algorithmic}[1]
    \Require
    A random forest $\mathcal{T}$;
    the index of testing sample $index$;
    \Ensure
    The testing result $O$.

    \For{$T_i\in RF$}
        \State \entity{CS} computes $r_i = $\protocol{FT-Testing}$(T_i, 1, index)$ and $\mathcal{R} = \mathcal{R}\cup r_i$
    \EndFor

    \If{$task$ is $0$} \# regression task
        \State \entity{CS} computes the final output $O = \frac{1}{|\mathcal{R}|}\sum_{i = 1}^{|\mathcal{R}|} r_i$
    \Else \xspace \# classification task
        \State \entity{CS} counts $r_i \in \mathcal{R}$, and selects the one with most votes as the final classification result $O$.
    \EndIf

    \State \entity{CS} returns $O$ to $u_0$.



  \end{algorithmic}
\end{algorithm}

\begin{algorithm}[ht]
  \caption{Federated Tree Prediction for Testing Data (\protocol{FT-Test})}
  \label{pro_ft_testing}
  \begin{algorithmic}[1]
    \Require
    A tree node $Node$;
    the index of testing sample $index$;
    current tree depth $d_c$;
    \Ensure
    The prediction result of a tree $r$.

    \If{$d_c \geq d_{max}$}
        \State \Return the weight of $Node$.
    \EndIf

    \State \entity{CS} extracts the split $[\![s_{\tau}]\!]_{pk_{\tau}}$ and corresponding split provider $u_{\tau}$ of current node of $T_i$.

    \State \entity{CS} re-encrypts the split $[\![s_{\tau}]\!]_{pk_{\Sigma}} = $\func{HReEnc}$(pk_{CS}, [\![s_{\tau}]\!]_{pk_{\tau}})$ and sends $[\![s_{\tau}]\!]_{pk_{\Sigma}}$ to $u_{\tau}$.

    \State $u_{\tau}$ computes $[\![s_{\tau}]\!]_{pk_{CS}} = $\func{ParHDec1}$(sk_{\tau},  [\![s_{\tau}]\!]_{pk_{\Sigma}})$, 
    encrypts the local feature value $[\![x_{index}]\!]_{pk_{\tau}} = $\func{HoEnc}$(pk_{\tau}, x_{index})$ 
    and invokes $[\![l]\!]_{pk_{\Sigma}} = $\func{HoLT}$([\![x_{\tau}]\!]_{\tau}, [\![s_{\tau}]\!]_{pk_{CS}})$.

    \State $u_{\tau}$ returns $[\![l]\!]_{pk_{CS}} = $\func{ParHDec1}$(sk_{\tau}, [\![l]\!]_{pk_{\Sigma}})$.

    \State \entity{CS} decrypts $l = $\func{ParHDec2}$(sk_{CS}, [\![l]\!]_{pk_{CS}})$ and updates $d_c = d_c + 1$.
    If $l$ is $0$, operate \protocol{FT-Test}$(Node\to rightChild,d_c, [\![\vec{x}]\!]_{pk_0})$; otherwise, \protocol{FT-Test}$(Node\to leftChild, d_c, [\![\vec{x}]\!]_{pk_0})$.


  \end{algorithmic}
\end{algorithm} 

\section{Security of Secure RF Prediction}\label{appendix_sec_prediction}
The security of secure RF prediction in a similar way to secure RF construction.

\begin{theorem}
\label{thm_predict_adv_ud}
    For secure RF prediction, there exists a PPT simulator $\xi_{UD}$ that can simulate an ideal view, which is computationally indistinguishable from the real view of $\mathcal{A}_{UD}$.
\end{theorem}

\noindent \textit{proof.}
\quad
    Two kinds of participants in \entity{UD} have to be simulated in this stage, the prediction requester and the normal participants.
    For corrupted participants, $\xi_{UD}$ can still use their local data to complete the protocol steps.
    For an honest prediction requester, $\xi_{UD}$ simulates it by using a randomly generated dummy request and encrypting the request with \func{HoEnc}.
    Base on Lemma~\ref{lem_enc_security}, the ideal encrypted request is computationally indistinguishable from a real one.
    For an honest normal participant, it only has to return a partially decrypted result to \entity{CS}.
    This can be simply simulated by asking $\mathcal{F}_t$ to operate \func{ParHDec1}, whose security is proved in Lemma~\ref{lem_enc_security}, and returns the output to $\mathcal{A}_{UD}$.
    In consequence, it is concluded that the simulated view is indistinguishable from the real view of $\mathcal{A}_{UD}$.
\QEDB

\begin{theorem}
\label{thm_predict_adv_cs}
    For secure RF prediction, there exists a PPT simulator $\xi_{CS}$ that can simulate an ideal view, which is computationally indistinguishable from the real view of $\mathcal{A}_{CS}$.
\end{theorem}

\noindent \textit{proof.}
\quad
    $\xi_{CS}$ can simulate the prediction stage in two steps.
    First, $\xi_{CS}$ combines the simulator in Theorem~\ref{thm_train_adv_server} to get the input of \func{HoLT}, i.e., the encrypted split.
    Second, $\xi_{CS}$ completes the iterative invocation of \protocol{FT-Predict} by running \func{ParHDec2} and outputs the final result.
    The homomorphic encryption keys in this stage are the same as Theorem~\ref{thm_train_adv_server}.
    From Lemma~\ref{lem_enc_security}, the two functions can be securely conducted.
    Therefore, the simulated view is indistinguishable from the real view of $\mathcal{A}_{CS}$.
\QEDB

\begin{theorem}
\label{thm_predict_adv_cc}
    For secure RF prediction, there exists a PPT simulator $\xi_{CC}$ that can simulate an ideal view, which is computationally indistinguishable from the real view of $\mathcal{A}_{CC}$.
\end{theorem}

\noindent \textit{proof.}
\quad
    In \sysname, the only task of \entity{CC} is to assist \entity{CS} to complete the computation of some homomorphic encryption based functions (referring to Appendix~\ref{appendix_functions}).
    Therefore, the security proof of \entity{CC} is totally based on Lemma~\ref{lem_enc_security}.
    Since Lemma~\ref{lem_enc_security} has been proved to be secure \cite{bresson2003simple}, there must be a simulator that can perfectly simulate $\xi_{CC}$. 
    The interested readers can refer to \cite{liu2016efficient} for detailed proof.
\QEDB

Based on the above theorems, we can simply derive that the RF prediction stage of \sysname is secure with different kinds of adversaries defined in our security model.

\section{Experiments of \sysname}
\subsection{Dataset Information}\label{appendix_dataset}
Table~\ref{table_dataset} shows the datasets used in our experiments.
\vspace{-4pt}
\begin{table}[!htbp]
\centering
\caption{Dataset Information}
\begin{tabular}{|c|p{36pt}|p{31pt}|c|}

\hline
Name & No. of Instances & No. of Features & Task Type \\


\hline
Adult Income & 48842 & 14 & Cla. \\

\hline
Bank Market & 45211 & 17 & Cla. \\

\hline
Drug Consumption & 1885 & 32 & Cla. \\

\hline
Wine Quality & 4898 & 12 & Cla. \\

\hline
Super Conduct & 21263 & 81 & Reg. \\


\hline
Appliance Energy & 19735 & 29 & Reg. \\

\hline
Insurance Company & 9000 & 86 & Reg. \\

\hline
News Popularity & 39797 & 61 & Reg. \\

\hline
\end{tabular}
\label{table_dataset}
\end{table}

\subsection{Evaluation Indicators}\label{appendix_indicators}
The computations of the six indicators are listed as follows.
\begin{gather*}
    ACC = \frac{TP + TN}{TP + TN + FP + FN}, \\
    RR = \frac{TP}{TP + FN}, \quad
    F1 = \frac{2TP}{2TP + FP + FN},
\end{gather*}
where $TP$, $FP$, $TN$, $FN$ are the numbers of true positive, false positive, true negative and false negative samples, respectively \cite{xu2015comprehensive}.

\begin{gather*}
    MSE = \frac{1}{n}\sum_{i = 1}^{n}(y_i - \hat{y}_i)^2,
    MAE = \frac{1}{n}\sum_{i = 1}^{n}|y_i - y_i|,\\
    R2 = 1 - \sum_{i = 1}^{n} (y_i - \hat{y}_i)^2 / \sum_{i = 1}^{n} (\bar{y}_i - \hat{y}_i)^2,
\end{gather*}
where $n$ is the number of validation samples, $y_i$ is the ground truth, $\hat{y}_i$ is the prediction result and $\bar{y}$ is the average of predicted results.

\end{document}